%% file: main.tex
\documentclass[journal,comsoc]{IEEEtran}

\usepackage[T1]{fontenc} 
\usepackage{setspace}

\usepackage[caption=false]{subfig}
\usepackage[pdftex]{graphicx}

\usepackage{tikz}

\definecolor{gold}{rgb}{0.85,.66,0}

\usepackage{cite}

\usepackage{amsmath}
\usepackage[cmintegrals]{newtxmath}

\usepackage{bm}

\makeatletter
\def\munderbar#1{\underline{\sbox\tw@{$#1$}\dp\tw@\z@\box\tw@}}
\makeatother

\usepackage{soul}
\usepackage[normalem]{ulem} 

\usepackage{algorithm}
\usepackage{algpseudocode}

\ifCLASSINFOpdf
\else
\fi

\newcommand{\brm}[1]{\mathbf{ #1 }}

\newcommand{\ltwonorm}[1]{\lVert #1 \rVert_{2}}
\newcommand{\fronorm}[1]{\lVert #1 \rVert_{\text{F}}}

\newcommand{\zeros}{\brm{0}}
\newcommand{\eye}[1]{\brm{I}_{#1}}

\newcommand{\Hhat}{\brm{H}}

\newcommand{\rss}{\operatorname{RSS}}

\newcommand{\diag}[1]{\operatorname{diag}( #1 )}
\newcommand{\trace}[1]{\operatorname{tr}( #1 )}

\DeclareMathOperator*{\argmax}{arg\,max}
\DeclareMathOperator*{\argmin}{arg\,min}

\begin{document}



\newcommand{\papertitle}{
Accelerated Randomized Methods for Receiver Design in Extra-Large Scale MIMO Arrays
}
\title{\papertitle}

\author{
{Victor Croisfelt}, 
{Abolfazl Amiri},
{Taufik Abrão},
{Elisabeth de Carvalho}, 
{Petar Popovski}
\thanks{This work was supported in part by the National Council for Scientific and Technological Development (CNPq) of Brazil, Grant 310681/2019-7, in part by the Coordenação de Aperfeiçoamento de Pessoal de Nível Superior, Brazil (CAPES), Financial Code 001 (88887.461434/2019-00), in part by the CONFAP-ERC Agreement H2020, Brazil, and in part by the Danish Council for Independent Research DFF-701700271.
} 
\thanks{V. Croisfelt is with the Electrical Engineering Department, Universidade de S\~{a}o Paulo, Escola Politécnica, São Paulo, Brazil; victorcroisfelt@usp.br}
\thanks{A. Amiri, E. de Carvalho, and P. Popovski are with the Department of Electronic Systems, Technical Faculty of IT and Design; Aalborg University,	Denmark; \{aba; edc; petarp\}@es.aau.dk.}
\thanks{T. Abrão is with the Electrical Engineering Department, State University of Londrina, PR, Brazil; taufik@uel.br.}%
}


\maketitle

\begin{abstract} 
Massive multiple-input-multiple-output (M-MIMO) features a capability for spatial multiplexing of large number of users. This number becomes even more extreme in extra-large (XL-MIMO), a variant of M-MIMO where the antenna array is of very large size. Yet, the problem of signal processing complexity in M-MIMO is further exacerbated by the XL size of the array. The basic processing problem boils down to a sparse system of linear equations that can be addressed by the randomized Kaczmarz (RK) algorithm. This algorithm has recently been applied to devise low-complexity M-MIMO receivers; however, it is limited by the fact that certain configurations of the linear equations may significantly deteriorate the performance of the RK algorithm. In this paper, we embrace the interest in accelerated RK algorithms and introduce three new RK-based low-complexity receiver designs. In our experiments, our methods are not only able to overcome the previous scheme, but they are more robust against inter-user interference (IUI) and sparse channel matrices arising in the XL-MIMO regime. In addition, we show that the RK-based schemes use a mechanism similar to that used by successive interference cancellation (SIC) receivers to approximate the regularized zero-forcing (RZF) scheme.
\end{abstract}

\begin{IEEEkeywords}
massive MIMO; extra-large scale massive MIMO; randomized Kaczmarz algorithm; receiver design.
\end{IEEEkeywords}

%
\IEEEpeerreviewmaketitle

\section{Introduction}\label{sec:intro}
Early deployments of fifth generation (5G) networks are already exploiting massive multiple-input multiple-output (M-MIMO) technology to cope with the rapid growth in the number of users and data traffic \cite{Bjornson2019}. The benefits from the M-MIMO topology come from the spatial multiplexing of the users on the same time-frequency resources. However, the common choice for compact antenna arrays limits the spatial dimension of such systems, reducing the performance gains achievable in practice. One way to enhance the promised benefits of M-MIMO is to scale up the number of antenna elements at the base station (BS). Systems that embrace antenna arrays of extremely large dimensions can better separate a large number of users, significantly increasing overall performance. This uncovers a new regime of M-MIMO referred to as the extra-large scale MIMO (XL-MIMO) \cite{Carvalho2020}. 

Despite the potential benefits, a disadvantage of extremely large antenna arrays is the excessively high computational complexity concerning signal processing at the receiver. The reason is that inter-user interference (IUI) management is necessary to deal with a large number of users, motivating the use of more intricate receiver designs. The canonical regularized zero-forcing (RZF) is one of these schemes that can offer near-optimum performance in many scenarios \cite{Bjornson2017c}. Unfortunately, applying the RZF scheme implies calculating the inverse of large matrices, which needs very high computational capacity at the processing units. This motivates the design of schemes that match the RZF performance, while offering complexity that scales better with the number of antennas and users.

Another practical challenge for receivers when increasing the dimension of the antenna arrays is the emergence of new channel effects. With a larger array, different users experience the same channel paths with variable energy or totally different channel paths. This effect results in a variable mean energy value along the array that is called \emph{spatial non-stationarities} \cite{Carvalho2020}. In contrast, the term \emph{spatial stationarity} refers to the case where the energy variations along the array is negligible. Non-stationarities give rise to sparse channel matrices due to the possibility that the user's energy is concentrated only in a small part of the large antenna array. This uneven energy distribution limits the performance of conventional linear receivers, \textit{e.g.}, zero-forcing (ZF) \cite{Ali2019a}. Thus, there is a need for low-complexity receiver designs that are aware of such non-stationarities.

\subsection{Related work}
Many recent works address the design of low-complexity receivers in the context of multi-antenna systems. One of the most common techniques consists of approximating the matrix inverse in the RZF scheme. There are three main approximation techniques: approximate matrix inversion algorithms \cite{Sessler2005,Kammoun2014}, matrix gradient search methods \cite{Yin2014}, and iterative solvers of systems of linear equations (SLEs) \cite{Gao2014,Dai2015,Boroujerdi2018a}. These methods provide ways to manage the performance and complexity trade-off. {However, they face some challenges that can decrease their applicability.} {The first two have limited control over the performance-complexity} management and can involve steps {that can still be} considered complicated and costly {from implementation point of view}. For example, the truncated polynomial expansion (TPE) technique used in \cite{Sessler2005,Kammoun2014} has iterations comprised of matrix products and further processing is needed to fine tune parameters. {Iterative solvers of SLEs, on the other hand, depend dramatically on their convergence rate. In this paper, we focus on the third category in order to increase its applicability using acceleration techniques that are premised on simplicity.}

Among the iterative solvers of SLEs, {the Kaczmarz algorithm \cite{Kaczmarz1937}} is a popular approach for solving very large SLEs, fitting well with our application scenario. In \cite{Strohmer2009a}, a randomized Kaczmarz (RK) algorithm was introduced and shown to have an excellent convergence behavior. The authors of \cite{Boroujerdi2018a} introduced a low-complexity receiver design to approximate the RZF scheme based on the RK algorithm of \cite{Strohmer2009a} for M-MIMO. There are two main features in favor of the RK-based RZF scheme of \cite{Boroujerdi2018a}. First, the scheme is \textit{simple}, meaning that there is no need to adjust any parameters other than the number of iterations or to know second-order channel statistics. Second, the scheme is \textit{flexible}, which means that it can easily control performance and complexity with great granularity by adjusting the number of iterations.

However, two known problems with the RK algorithm were not treated in \cite{Boroujerdi2018a}. The RK algorithm randomly selects one of the SLE equations to be solved in a given iteration. This equation sampling is based on a {probability criterion} where probabilities are proportional to the energy of the equations, giving rise to the following weaknesses: (a) low-energy equations are rarely selected, and (b) performance of the RK algorithm is deteriorated when the energy of the equations are very similar \cite{Bai2018a,Censor2009,Strohmer2009}. We refer to these weaknesses as the \textit{problem with rare equations} and the \textit{curse of uniform normalization}, respectively. Then, the low-complexity receiver of \cite{Boroujerdi2018a} performs poorly when some users are located at the cell-edge or a user power control scheme has been employed. Because of this, we call the receiver scheme in \cite{Boroujerdi2018a} as nRK, short for naive RK. Besides the above intrinsic issues, it has been found in \cite{Bai2018a} that the \textit{RK algorithm can fail under certain sparsity conditions}, hindering the operation of the nRK in sparse channels characteristics of the XL-MIMO regime.

Recent interest in solving sparse SLEs for neural network training and other machine learning problems has motivated the research on accelerated RK algorithms, such as the greedy RK (GRK) of \cite{Bai2018a} and the randomized sampling Kaczmarz (RSK) of \cite{Sun2020}. The accelerated RK algorithms can address the three presented weaknesses of the RK algorithm to some extent. In this paper, we embrace this observation and introduce three different accelerated RK-based receiver designs to compete against the nRK scheme of \cite{Boroujerdi2018a}.

Distributed receiver designs are also being studied to further alleviate the complexity in the XL-MIMO systems \cite{Amiri2020,Wang2020}. We share the belief that the distributed approach is the way to further reduce complexity when it comes to XL-MIMO systems, due to the excess of complexity and information management brought by the large number of antennas. For the sake of tractability, however, here we focus on a centralized receiver design in order to cover simultaneously the discussion of both M-MIMO and XL-MIMO regimes. Since centralized designs do not suffer from the inevitable performance loss given the decentralization process \cite{Amiri2020}, they can be advantageous when the number of antenna modules is limited and the hardware does not suffer from unsustainable processing capacity and excessive information communication. Furthermore the distributed framework derived in our previous work \cite{Rodrigues2020} can be used to generalize the receivers presented herein for a XL-MIMO system comprised of sub-arrays \cite{Carvalho2020}.

\subsection{Contributions}
In this paper, we introduce three low-complexity receiver designs based on the accelerated RK-algorithms for M-MIMO and XL-MIMO systems. These acceleration techniques are using the following heuristics: (i) the sampling without replacement (SwoR) technique, (ii) the GRK algorithm of \cite{Bai2018a}, and (iii) the RSK algorithm of \cite{Sun2020}. To the best of our knowledge, this is the first work that uses these accelerated RK-based algorithms to design receivers for multi-antenna systems. Our schemes work by approximating the performance of the RZF, while providing control over its complexity. This control is realized by only adjusting the number of iterations of the algorithms. The proposed schemes are executed at a central processing unit (CPU).  Below, we summarize the contributions of this work:
\begin{itemize}
    \item We present three flexible receivers that are able to select points of operation from a two-dimensional space defined by performance and complexity. The upper bound of performance is provided by the RZF scheme and the performance range is discretized by the number of iterations.
    \item We show that one can interpret the RK-based receivers in general to perform a kind of successive interference cancellation (SIC) procedure, giving us a better understanding of how the RK algorithms work when approaching the RZF scheme from the standpoint of classical literature.
    \item We provide a detailed complexity analysis, showing that our schemes have more scalable computational complexities with respect to the number of antennas and users.
\end{itemize}

The remainder of this paper is organized as follows. Section \ref{sec:systemmodel} defines a single-cell uplink system model suitable for the M-MIMO and XL-MIMO regimes. Section \ref{sec:preliminaries} introduces the mathematical framework needed to approximate the RZF scheme based on the RK algorithms. The proposed accelerated RK-based RZF schemes are described in Section \ref{sec:accrk}, while Section \ref{sec:interpret} provides a better interpretation of them. Complexity analysis and numerical results are given in Section \ref{sec:cc-num-res} followed by the main conclusions summarized in Section \ref{sec:conclusion}.

\noindent \textit{Notations:} We use upper and lower case boldface to denote matrices and vectors, while non-boldface are used for constants. Discrete and continuous sets are given by calligraphic $\mathcal{X}$ and blackboard bold $\mathbb{X}$. Cardinality of a set is given by $|\mathcal{X}|$. The $n$-th element of $\brm{x}$ is denoted as $x_n$. The $m,n$-th element of the matrix $\brm{X}$ is $[\brm{X}]_{m,n}$, while $[\brm{X}]_{:,n}$ represents the $n$-th column vector of $\brm{X}$. Vertical and horizontal matrix concatenations are $[\brm{X};\brm{Y}]$ and $[\brm{X},\brm{Y}]$, respectively. We indicate transpose and Hermitian transpose by $(\cdot)^\transp$ and $(\cdot)^\htransp$. Identity matrix of size $n$ is denoted as $\eye{n}$, while $\brm{0}_{m\times n}$ is an $m\times{n}$ matrix of zeros. Trace and diagonal matrix operators are denoted respectively by $\trace{\cdot}$ and $\diag{\cdot}$. The $l2$- and Frobenius norms are given by $\ltwonorm{\brm{x}}$ and $\fronorm{\brm{X}}$, respectively. Gaussian distribution is represented by $\mathcal{N}(\cdot,\cdot)$, whereas circularly symmetric complex-Gaussian distribution is $\mathcal{CN}(\cdot,\cdot)$. The indicator function with argument $x$ over the set $\mathcal{A}$ is denoted as $\chi_{\mathcal{A}}(x)$, where $\chi_{\mathcal{A}}(x)=1$ if $x\in\mathcal{A}$, and zero otherwise. Floor operation is $\lfloor\cdot\rfloor$.

\section{System Model}\label{sec:systemmodel}
We consider the uplink payload data transmission of an M-MIMO system wherein a BS with $M$ antennas simultaneously serves a total of $K<M$ single-antenna users. For convenience, the group of users is indexed by the set of integers $\mathcal{K}=\{1,2,\dots,K\}$, while $\mathcal{M}=\{1,2,\dots,M\}$ is the set of antenna indexes. {Moreover,} we assume the block-fading channel model where the {channel vector $\brm{h}_k\in\mathbb{C}^{M\times 1} $ of user ${k\in\mathcal{K}}$ is constant and frequency-flat} within a coherence block \cite{Bjornson2017c}. {When all users transmit simultaneously, the BS receives the following narrowband baseband signal $\brm{y}\in\mathbb{C}^{M\times 1}$:}
\begin{equation}
    \brm{y}=\sqrt{\rho}\brm{H}\brm{x} + \brm{n},%
    \label{eq:receivedsignal}%
\end{equation}
where $\rho$ is the user transmit power, $\brm{H}\in\mathbb{C}^{M\times{K}}=[\brm{h}_{1},\brm{h}_2,\dots,\brm{h}_K]$ is the channel matrix perfectly known by the BS, $\brm{x}\in\mathbb{C}^{K\times 1}$ is the transmitted signal vector, and $\brm{n}\in\mathbb{C}^{M\times 1}\sim{\mathcal{CN}}(\brm{0},\sigma^{2}\eye{K})$ is the receiver noise vector {with noise power $\sigma^2$.} The vector $\brm{x}$ is composed of the modulated transmission symbols sent by each user independently, where $x_k$ is drawn from a normalized constellation sequence $\mathcal{X}$. Our goal is to design an efficient and reliable receiver that coherently combines the $M$ observations of the received signal $\brm{y}$ and produce a \textit{soft estimate} $\hat{\brm{x}}$ for $\brm{x}$. Throughout this work, we consider a traditional baseband processing architecture, where a CPU is responsible for all processing activities related to the signal reception in the antenna array. 

\subsection{General Channel Model}\label{subsec:chnmodel}
We adopt the correlated and non-stationary Rayleigh fading channel model proposed in \cite{Ali2019a}, {suitable for transmissions at sub-6 GHz frequencies.} Following this model, the channel vector of user $k\in\mathcal{K}$ is defined by $\brm{h}_k\sim{\mathcal{CN}}(\brm{0},\brm{\Theta}_k)$, where $\brm{\Theta}_k\in\mathbb{C}^{M\times{M}}$ is the general channel covariance matrix. This matrix can be decomposed as \cite{Ali2019a}:
\begin{equation}
    \brm{\Theta}_k=\brm{D}^{\frac{1}{2}}_{k}\brm{R}_k\brm{D}^{\frac{1}{2}}_{k}.%
    \label{eq:generalcovmatrix}
\end{equation}
where $\brm{R}_{k}\in\mathbb{C}^{M\times{M}}$ is the spatial correlation matrix and $\brm{D}_{k}\in\{0,1\}^{M\times M}$ is an indicator diagonal matrix. Further, the vector with the large-scale coefficients of user $k$ is defined as $\boldsymbol{\beta}_k=\diag{\brm{R}_k}=[\beta^{1}_k,\dots,\beta^{M}_k]^\transp$. The diagonal matrix $\brm{D}_{k}$ models the portions of the antenna array "seen" by each user through the concept of visibility regions (VRs) \cite{Carvalho2020}, where $[\brm{D}_{k}]_{m,m}=1$ indicates that antenna $m\in\mathcal{M}$ sees user $k\in\mathcal{K}$, $[\brm{D}_{k}]_{m,m}=0$ indicates otherwise. For convenience, we assume that $\trace{\brm{D}_k}=D, \ \forall k\in\mathcal{K}$, where $D\leq M$ is the number of \textit{visible} antennas. The term visible indicates that only $D$ antennas have the major contribution in the communication of any user to the BS, but the visible antenna indices can differ between users. One of the key differences between the M-MIMO and XL-MIMO regimes is in their corresponding $\brm{D}$ matrix that determines the stationarity of the received energy over the BS array. For the stationary case $\brm{D}=\eye{M}$, since the array is compact. Spatial non-stationarities impose a sparse structure into the channel matrix $\brm{H}$ that can be exploited for simpler receiver designs. In fact, one of the main motivations of this work is to design efficient receivers using such information to reduce the computational complexity.

\begin{table*}[htp]
    \centering
    \caption{Summary of the Low-Complexity Accelerated RK-Based RZF Receiver Designs}
    \label{tab:summary}
    \resizebox{\textwidth}{!}{%
    \begin{tabular}{|c|c|c|c|}
        \hline
        \textbf{Scheme} & \textbf{Acceleration Method} & \textbf{Advantages} & \textbf{Disadvantages} \\ \hline
        nRK-RZF \cite{Boroujerdi2018a} & none & \begin{tabular}[c]{@{}c@{}}least costly iteration\\ $\downarrow$ complexity due to sparsity (XL-MIMO)$^\dagger$\end{tabular} & \begin{tabular}[c]{@{}c@{}}$\downarrow$ performance for cell-edge users and user power control\\$\downarrow$ weak against IUI and sparsity\end{tabular} \\ \hline
        \begin{tabular}[c]{@{}c@{}}RK-RZF\\(Algorithm 1)\end{tabular} & SwoR & \begin{tabular}[c]{@{}c@{}}\textbf{best benefit-cost ratio}\\$\uparrow$ performance for cell-edge users and user power control\\$\uparrow$ robust to sparsity than nRK-RZF\end{tabular} & \begin{tabular}[c]{@{}c@{}}$\pm$ robust against IUI\\iteration cost grows linearly with $K$\end{tabular} \\ \hline
        \begin{tabular}[c]{@{}c@{}}GRK-RZF\\(Algorithm 2)\end{tabular} & complete residual info. & \begin{tabular}[c]{@{}c@{}}\textbf{best under extreme conditions}\\$\uparrow$ performance for cell-edge users and user power control\\$\uparrow$ robust against IUI and sparsity\\smallest number of iterations to converge\end{tabular} & most costly iterations \\ \hline
        \begin{tabular}[c]{@{}c@{}}RSK-RZF\\(Algorithm 3)\end{tabular}  & partial residual info. & \begin{tabular}[c]{@{}c@{}}same from GRK-RZF to a much smaller extent\\ intermediate iteration cost (between RK and GRK)\end{tabular} & $\uparrow$ number of iterations to converge w.r.t. GRK \\ \hline
    \end{tabular}%
    }
    \flushleft {\footnotesize $^\dagger$ All other receivers inherent the complexity reduction due to sparsity.}
\end{table*}

\section{Preliminaries}\label{sec:preliminaries}
In this section, we introduce the RZF scheme as one of the state-of-the-art receivers used in the literature \cite{Bjornson2017c}. We argue that the straightforward implementation of the RZF scheme may not be attractive from the hardware point of view. To solve this problem, we interpret its solution as an optimization problem that can be solved through an SLE. Then, we describe the process of acquiring a consistent SLE that meets our needs. 

A conventional scheme suitable for a scenario where IUI is a problem and the signal-to-noise-ratio (SNR) of users may vary overly is the RZF scheme \cite{Bjornson2017c}. Employing the RZF to combine coherently the payload information in $\brm{y}$ yields in \cite{Bjornson2017c}%
\begin{equation}
    \label{eq:zf_rzf}
    \brm{\hat{x}}^{\text{RZF}}=\left(\brm{V}^{\text{RZF}}\right)^{\htransp}\brm{y}=\left(\brm{H}^{\htransp}\brm{H}+\xi\eye{K}\right)^{-1}\brm{H}^\htransp\brm{y},%
\end{equation}
where $\brm{V}^{\text{RZF}}\in\mathbb{C}^{M\times K}$ is the RZF receive combining matrix, $\brm{\hat{x}}^{\text{RZF}}\in\mathbb{C}^{K\times 1}$ is the RZF soft estimate, $\xi=\sigma^2/\rho$ is the inverse of the pre-processing user transmit SNR, $\brm{G}\in\mathbb{C}^{K\times K}=\Hhat^{\htransp}\Hhat$ is the channel Gramiam matrix, and $\brm{R}_{\brm{y}\brm{y}}\in\mathbb{C}^{K\times K}=\brm{G}+\xi\eye{K}$ is the sample covariance matrix of the received signal $\brm{y}$.

The classical RZF scheme can be viewed as the solution of the following optimization problem \cite{Boroujerdi2018a}:%
\begin{equation}
    \label{eq:optzprob}
    { \brm{w}^\star\,=\,\,\,}  \argmin_{\brm{w}\in\mathbb{C}^{K\times 1}}\ltwonorm{\brm{H}\brm{w}-\brm{y}}^2 + \xi\ltwonorm{\brm{w}}^2,%
\end{equation}
where $\brm{w}^\star$ is the optimal solution and corresponds to $\brm{\hat{x}}^{\text{RZF}}$. {The proof simply follows by taking the derivative of the $l_2$-regularized least-squares cost function above and equating it to zero.} A compact form of the cost function is $\ltwonorm{\brm{B}\brm{w}-\brm{y}_0}^2$, where $\brm{B}=[\brm{H};\sqrt{\xi}\eye{K}]\in\mathbb{C}^{(M+K)\times{K}}$ and $\brm{y}_0=[\brm{y};\zeros_{K\times 1}]\in\mathbb{C}^{(M+K)\times{1}}$. Naturally, the solution of this optimization problem can be obtained by solving the thin SLE $\brm{B}\brm{w}=\brm{y}_0$.

The presence of noisy observations in $\brm{y}$ hinders the use of iterative solvers over $\brm{B}\brm{w}=\brm{y}_0$. On the other hand, this SLE is inconsistent; meaning that the noisy observations in $\brm{y}$ make $\brm{y}_0$ not lie in the range of $\brm{B}$ \cite{Meyer2000}. Thus there is no solution set. It is preferable to obtain a consistent SLE with minimum additional complexity cost. We use the transformation proposed in \cite{Boroujerdi2018a} that yields the following consistent, fat SLE:%
\begin{equation}
    \label{eq:consistent_linearsys}%
    \brm{B}^\htransp\brm{z}=\brm{b}=\brm{H}^{\htransp}\brm{y},%
\end{equation}
where $\brm{z}\in\mathbb{C}^{(M+K)\times 1}=[\brm{u}\in\mathbb{C}^{M\times 1};\sqrt{\xi}\brm{v}\in\mathbb{C}^{K\times 1}]$. The minimum-norm solution to the SLE above is given by $\brm{u}=\Hhat\brm{\hat{x}}^{\text{RZF}}$ and $\brm{v}=\brm{\hat{x}}^{\text{RZF}}$. One can note that the $k$-th equation of this SLE can be associated with obtaining the $k$-th component of $\brm{v}$, which solution is $\hat{x}^{\text{RZF}}_k$ of user $k\in\mathcal{K}$. \emph{Hence, we can use the terms equation and user interchangeably.}

\noindent\textbf{Remark 1.} \textit{(MR Receiver).}
The vector with constant terms {$\brm{b=}\brm{H}^{\htransp}\brm{y}$} in \eqref{eq:consistent_linearsys} is the maximum-ratio (MR) soft estimate $\brm{\hat{x}}^{\mathrm{MR}}$ and the price to pay for consistency \cite{Bjornson2017c}. Therefore, the respective upper and lower bounds of receiver performance and complexity are given by the MR scheme in this work. 

\noindent\textbf{Remark 2.} \textit{(Normal Equations).} 
The authors of \cite{Wu2020} designed Kacmarz-based receivers using the normal SLE {$\brm{B}^\htransp\brm{B}\brm{w}=\brm{B}^\htransp\brm{y}_0$}. The {solutions} discussed here are easily extended to this case as well. Here, we use the SLE in \eqref{eq:consistent_linearsys} to avoid the additional complexity of acquiring the normal SLE.

\noindent\textbf{Remark 3.}{ (\textit{Three Challenges}).}
(i) $\brm{B}^\htransp$ does not have symmetry properties, impeding the application of some classical iterative methods, \textit{e.g.}, conjugate gradient \cite{Press2007}. (ii) the amounts of calculations and storage are limited due to wireless nature, hindering the use of common acceleration techniques, such as preconditioning \cite{Press2007}.\footnote{It is common the use of a relaxation parameter to improve convergence of RK methods \cite{Bai2019}. However, this expedient requires adjusting the regularization parameter.} (iii) $\brm{B}^\htransp$ is a sparse full rank matrix when the channel is non-stationary $D\neq{M}$, making the SLE difficult to be solved by some methods, \textit{e.g.}, the RK algorithm \cite{Strohmer2009a} depending on the sparse structure. The methods proposed here aim to overcome these challenges.

\section{Low-Complexity Receiver Designs Based on Accelerated RK Algorithms}\label{sec:accrk}
In this section, we exploit recently established acceleration techniques for the RK algorithm to increase the applicability of RK-based receiver designs in both M-MIMO and XL-MIMO regimes. We start with an overview of the three introduced accelerated RK-based receivers, justifying the reasons for using the chosen methods and indicating the advantages and disadvantages of each. Then, we give a detailed presentation for each of the schemes. 

\subsection{Overview: Proposed Receivers}\label{subsec:overview}
We present three receivers based on variations of the RK algorithm: \textbf{(i) RK-RZF:} a receiver that improves the overall performance of nRK-RZF \cite{Boroujerdi2018a} using a SwoR-based acceleration technique, which is simpler than those used in the other two schemes; \textbf{(ii) GRK-RZF:} a greedy scheme that exploits the residual information of the SLE in \eqref{eq:consistent_linearsys} to further accelerate convergence \cite{Bai2018a}; \textbf{(iii) RSK-RZF:} a scheme introduced to deal with the complexity disadvantages of the GRK-RZF whilst still exploits part of the acceleration provided by the residual information \cite{Sun2020}. The "-RZF" suffix explicitly denotes that the performance of the RZF scheme is being emulated by the RK-based receivers.

Table \ref{tab:summary} summarizes the main differences and advantages/disadvantages of each new accelerated RK-based RZF schemes. The RK-RZF scheme has the best benefit-cost ratio among the three proposed receivers. This means that the RK-RZF is able to reduce the complexity of the RZF scheme with little performance losses for typical numbers of antennas $M$ and users $K$ of the M-MIMO and XL-MIMO systems. On the other hand, the performance of the RK-RZF is drastically affected by high levels of IUI and/or sparsity, and when operating at high SNR regime. The GRK-RZF scheme works better under these extreme conditions, being more robust against IUI, sparsity, and increased SNR. However, the price to pay for these gains can turn the GRK-RZF receiver very costly. The RSK-RZF scheme is an effort to reduce the cost while hold part of the benefits of the GRK-RZF. The region of applicability of the RSK-RZF receiver is very limited though.

\begin{figure}[!htbp]
    \vspace{-3mm}
    \centering
    \includegraphics[trim=2.5cm 0.0cm 0cm 0cm, clip=true, width=\linewidth]{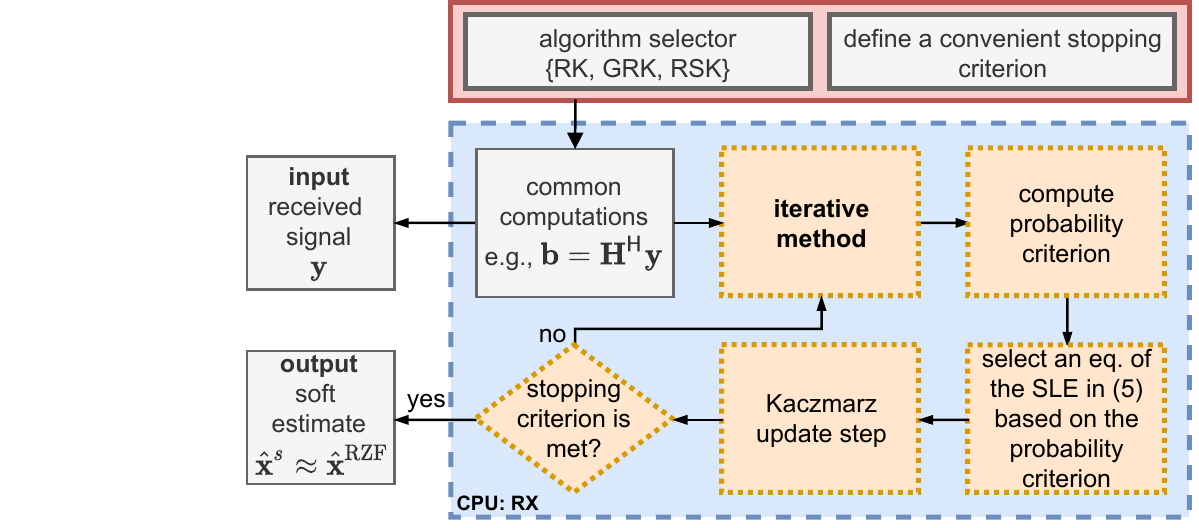}
    \vspace{-6mm}
    \caption{Illustration of the basic steps realized by the proposed low-complexity receivers based on accelerated RK algorithms in a centralized baseband architecture, where the CPU is carrying out the signal reception (RX).}
    \label{fig:basic-operation}
\end{figure}

Figure \ref{fig:basic-operation} illustrates the main steps that are common to all the proposed accelerated RK-based RZF receivers when considering a centralized baseband architecture coordinated by a CPU. First, the CPU uses the received signal $\mathbf{y}$ to calculate the common information which is fixed for all the iterations.\footnote{As a practical matter, if the coherence block is large enough, it is more efficient to calculate $\brm{V}^{\text{RZF}}$ in \eqref{eq:zf_rzf} just once and then use it until the end of the coherence block. Actually, the schemes described here can be generalized in this way by following the lines in \cite{Boroujerdi2018a,Rodrigues2019,Rodrigues2020}.} Then, depending on the selected scheme (from Table \ref{tab:summary}), it starts with the user symbol estimation process. In general, the probability criterion used to select the equations from the SLE in \eqref{eq:consistent_linearsys} differs between the algorithms, while the steps used to update the solution are the same. We refer to this set of steps as the \textit{Kaczmarz update step}, since it follows the classical Kaczmarz algorithm \cite{Kaczmarz1937}. If a pre-defined stopping criterion is fulfilled, the iterative method converges and the CPU obtains the soft estimate $\hat{\brm{x}}^{s}$, which is an approximation of $\hat{\brm{x}}^{\text{RZF}}$, where the superscript $s$ indexes the proposed schemes in $\{\text{RK}, \text{GRK}, \text{RSK}\}$ according to Table \ref{tab:summary}. Otherwise, the algorithm will continue until a certain maximum number of iterations. The complexity analysis and the stopping criterion are discussed in Subsection \ref{subsec:cc}.

\subsection{Randomized Kaczmarz Algorithm with SwoR}\label{subsec:rk:algo}
Algorithm \ref{algo:RK}\footnote{Two key observations about the description of all algorithms: (a) a count in terms of floating-point operation per second (FLOPs) is annotated after "\%" and (b) the keyword "Store" stands for one time calculations.} summarizes the RK method applied to solve \eqref{eq:consistent_linearsys} when adopting the SwoR technique in Step \ref{algo:RK:swor}. We refer to Algorithm \ref{algo:RK} as the RK-RZF scheme. {Except for the application of the SwoR technique, the description of the nRK-RZF \cite{Boroujerdi2018a} scheme is the same as that used in Algorithm \ref{algo:RK}. However, this seemingly small modification leads to important implications as we discuss below.}

The RK-RZF scheme works as follows: Steps 1-7 are comprised of initialization of variables and common computations that will be used throughout the iterative process. In Step 7, the sampling probability vector $\brm{p}$ is calculated in \eqref{eq:prob:energy}, representing the probability criterion used to select the equations from the SLE in \eqref{eq:consistent_linearsys}. As long as a stopping criterion is not met, the algorithm randomly selects in Step 9 one of the equations based on $\brm{p}$, where the superscript $(t)$ indicates the current iteration. After choosing an equation index $i^{(t)}\in\mathcal{K}$, the method projects orthogonally the last iterative solution $\brm{z}^{(t)}=[\brm{u}^{(t)},\brm{v}^{(t)}]$ onto the solution hyperplane $b_{i^{(t)}}=\brm{h}^\htransp_{i^{(t)}}\brm{u}^{(t)} + \xi v_{i^{(t)}}$, as described in Step 10. This orthogonal projection is seen as the residual $r^{(t)}_{i^{(t)}}$ in relation to the $i^{(t)}$-th equation. In Step 11, this residual is normalized by the energy of the chosen equation $\ltwonorm{\brm{h}_{i^{(t)}}}^2+\xi$. In Steps 12 and 13, the solution $\brm{z}^{(t)}$ is updated into $\brm{z}^{(t+1)}$, considering the contribution $\gamma^{(t)}$ of the normalized orthogonal projection of the last iterate over equation $i^{(t)}$. The \textit{Kaczmarz update step} in Fig. \ref{fig:basic-operation} can be defined according to the realization of Steps 10-14 and is common to all the algorithms in this paper. When the stopping criterion is met, the iterative process terminates and $\brm{v}^{(t)}$ is considered to be the RK-RZF soft estimate $\brm{\hat{x}}^{\text{RK}}$, an approximation of $\brm{\hat{x}}^{\text{RZF}}$.

\noindent{\textbf{SwoR.}} Let $\mathcal{P}^{(t)}$ denote the population from which the equations are sampled available in Step 9 at iteration $t$. At $t=0$, we have $\mathcal{P}^{(0)}=\mathcal{K}$. Applying the SwoR technique implies that $\mathcal{P}^{(t+1)}=\mathcal{P}^{(t)}\setminus\{i^{(t)}\}$ until the end of a sweep. At an arbitrary iteration $t'$, we define a \textit{sweep} as the cycle of $K$ iterations that consists of bringing $|\mathcal{P}^{(t')}|$ from $|\mathcal{K}|$ elements to $1$ element. After the end of a sweep, a new sweep begins with $\mathcal{P}^{(t'+K)}=\mathcal{K}$ and so on. {Because $\mathcal{P}^{(t)}$ changes at each new iteration, the sampling probability vector $\textbf{p}$ in Step \ref{algo:RK:prob} needs to be constantly re-scaled in Step \ref{algo:RK:swor} considering the elements in $\mathcal{P}^{(t)}$.}

\subsubsection{Probability criterion based on energy information of the equations}
The key feature of the RK algorithm \cite{Strohmer2009a} is the sub-optimal probability criterion in \eqref{eq:prob:energy} that dictates how the equations of the SLE in \eqref{eq:consistent_linearsys} will be selected.\footnote{Due to natural wireless channel variations, the SLE in \eqref{eq:consistent_linearsys} is constantly changing. Finding the optimum probability of sampling the equations is very costly and should be computed many times, and, therefore, avoided here.} This criterion is based on the \textit{energy information of the equations}. Note that the $k$-th entry of the sampling probability vector $\brm{p}$ is the ratio of the regularized channel gain of user $k$ to the sum of the {regularized} channel gains of all users. And that the desired solution $v^{(t)}_k$ is only updated if the $k$-th equation is selected. Two bad phenomena can occur if $\brm{p}$ is poorly scaled: (a) the weakest users will not be selected as often resulting in a poor performance; (b) convergence performance is naturally degraded if the users experience similar channel gains, which implies that $p_k\approx1/K, \ \forall k\in\mathcal{K}$. Both problems can be partly eliminated in a heuristic way by the SwoR technique in Step \ref{algo:RK:swor} of Algorithm \ref{algo:RK}. First, the SwoR technique avoids selecting the same equation in sequence, increasing the frequency of selection of the weakest users. Second, the SwoR is changing the selection probabilities constantly to lower the effect of uniform normalization curse. On the flip side, the SwoR technique comes with the price of re-scaling $\brm{p}$ {in Step \ref{algo:RK:swor}} after each new iteration, which causes the cost per iteration to grow linearly with $K$. Motivated by these limited solutions provided by the RK-RZF, in the following we seek other acceleration methods to improve the nRK-RZF \cite{Boroujerdi2018a} receiver.

\begin{algorithm} 
    \caption{{RK-with-SwoR-Based Receiver} (RK-RZF)\label{algo:RK}}
    \algrenewcommand{\algorithmiccomment}[1]{\hskip3em\% #1}
    \algrenewcommand\algorithmicrequire{\textbf{input:}}
    \algrenewcommand\algorithmicensure{\textbf{output:}}
    \begin{algorithmic}[1]
        \Require$\brm{H}$, $\brm{y}$, $M$, $K$, $\xi$%
        \Ensure$\brm{\hat{x}}^{\text{RK}}$, $T^{\text{RK}}$%
        \State
        $\brm{b}=\Hhat^\htransp\brm{y}$\qquad\qquad\quad\,\,\,\,\,\,{\%{ $8KM-2K$ FLOPs}}%
        \State Store $\{\ltwonorm{\brm{h}_k}^2+\xi\}$\qquad{\%{ $8KM-K$ FLOPs}}%
        \State Store $\fronorm{\brm{H}}^{2}+K\xi=\sum_{k\in\mathcal{K}}(\ltwonorm{\brm{h}_k}^2+\xi)$\qquad{\%{ $K-1$ FLOPs}}%
        \State $\brm{u}^{(0)}\in\mathbb{C}^{M\times 1}=\brm{0}_{M\times1}$%
        \State $\brm{v}^{(0)}\in\mathbb{C}^{K\times 1}=\brm{0}_{K\times1}$%
        \State $t=0$%
        \State Probab. criterion based on energy info. of eqs. $\brm{p}\in\mathbb{R}^{K\times 1}$: \label{algo:RK:prob}
        \begin{equation}
            p_k=\frac{\ltwonorm{\brm{h}_{k}}^2 + \xi}{\fronorm{\Hhat}^{2} + K\xi}, \ \forall k \in \mathcal{K}
            \label{eq:prob:energy}
        \end{equation}
        \While{$\text{stopping criterion}\text{\textbf{ is }}\mathrm{False}$}\label{algo:RK:termination}%
        \State pick $i^{(t)}\in\mathcal{K}$ by {re-scaling} $\brm{p}$ w/ SwoR\,\,{\%{ $K$ \small FLOPs}}\label{algo:RK:swor}%
        \Statex \textbf{Kaczmarz update step (Steps 10-14):}
        \State \( r^{(t)}_{i^{(t)}}= b_{i^{(t)}}-\brm{h}^\htransp_{i^{(t)}}\brm{u}^{(t)} - \xi v^{(t)}_{i^{(t)}} \)\qquad{\%{ $8M+4$ FLOPs}}%
        \State \(\gamma^{(t)}= {r^{(t)}_{i^{(t)}}}/({\ltwonorm{\brm{h}_{i^{(t)}}}^2 + \xi})\)\qquad\quad\,\,\,\,\,{\%{ $2$ FLOPs}}%
        \State $\brm{u}^{(t+1)}=\brm{u}^{(t)}+\gamma^{(t)}\brm{h}_{i^{(t)}}$\qquad\qquad\,\,\,\,{\%{ $8M$ FLOPs}}%
        \State $\brm{v}^{(t+1)}=\brm{v}^{(t)}+\gamma^{(t)}[\eye{K}]_{:,i^{(t)}}$\qquad\quad\,{\%{ $2$ FLOPs}}%
        \State $t=t+1$%
        \EndWhile%
        \State $\brm{\hat{x}}^{\text{RK}}=\brm{v}^{(t)}$, $T^{\text{RK}}=t$
    \end{algorithmic}
\end{algorithm}

\subsection{Greedy Randomized Kaczmarz Algorithm}
The GRK algorithm is {an accelerated version of the RK algorithm proposed in \cite{Bai2018a}}. The main idea is to eliminate the equations with larger residuals as quickly as possible. This heuristic design deals with the problem of rare equations and the curse of uniform normalization present in the RK algorithm. Beyond that, the GRK has improved convergence in comparison to the RK algorithm when solving sparse SLEs in general. The application of the GRK to solve \eqref{eq:consistent_linearsys} is given in Algorithm \ref{algo:GK}, namely, the GRK-RZF scheme. 

\begin{algorithm}
    \caption{GRK-Based Receiver (GRK-RZF)\label{algo:GK}}
    \algrenewcommand{\algorithmiccomment}[1]{\hskip3em\% #1}
    \algrenewcommand\algorithmicrequire{\textbf{input:}}
    \algrenewcommand\algorithmicensure{\textbf{output:}}
    \begin{algorithmic}[1]
        \Require{$\brm{H}$, $\brm{y}$, $M$, $K$, $\xi$}%
        \Ensure$\brm{\hat{x}}^{\text{GRK}}$, $T^{\text{GRK}}$%
        \State Repeat Steps 1-6 of Algorithm \ref{algo:RK}{\%{ $16KM-2K-1$ FLOPs}}%
        \State Store $(\fronorm{\brm{H}}^{2}+K\xi)^{-1}$\qquad{\%{ $1$ flop}}%
        \While{$\text{stopping criterion}\text{\textbf{ is }}\mathrm{False}$}\label{algo:GRK:termination}%
        \State $\brm{r}^{(t)}\in\mathbb{C}^{K\times{1}}$ w/\qquad{\% $8KM-8$ FLOPs}\label{algo:GRK:res}%
        $$
        r^{(t)}_{k}=b_k-\brm{h}_{k}^\htransp \brm{u}^{(t)} - \xi v^{(t)}_k%
        $$
        \State $\vv{\mathrm{SAR}}^{(t)}\in\mathbb{R}^{K\times 1} \text{ with } \mathrm{SAR}^{(t)}_{k} = \lvert r^{(t)}_{k}\rvert^2$\quad{\%{ $3K$ FLOPs}}\label{algo:GRK:uk01}%
        \State $\rss^{(t)}=\sum_{k\in\mathcal{K}}\mathrm{SAR}^{(t)}_{k}$\quad\,\,{\%{ $K-1$ FLOPs}}%
        \State Compute $\epsilon^{(t)}$ as\quad{\%{ $2K+3$ FLOPs}}\label{algo:GRK:uk02}
        $$
        \epsilon^{(t)}=\frac{1}{2}\left(\frac{1}{\rss^{(t)}}\max_{j\in\mathcal{K}}\left\{\frac{\mathrm{SAR}^{(t)}_{j}}{\ltwonorm{\brm{h}_j}^2 + \xi}\right\}+\frac{1}{\fronorm{\brm{H}}^{2} + K\xi} \right)%
        $$
        \State Get the set of working equations:\qquad{\%{ $K+1$ FLOPs}}\label{algo:GRK:working}
        $$
        \mathcal{U}_t=\left\{k : \mathrm{SAR}^{(t)}_{k} \geq \epsilon^{(t)}\rss^{(t)}\left(\ltwonorm{\brm{h}_k}^2 + \xi\right)\right\}%
        $$
        \State Probab. criterion based on complete residual info. \(\brm{p}^{(t)}\in\mathbb{R}^{K\times 1}\): \qquad{\%{ $K$ FLOPs}}\label{algo:GRK:probability}
        \begin{equation}
        \setlength{\nulldelimiterspace}{0pt}
        p^{(t)}_k=\left\{\begin{IEEEeqnarraybox}[\relax][c]{l's}
        \frac{\mathrm{SAR}^{(t)}_k}{\sum_{j\in\mathcal{U}_t}\mathrm{SAR}^{(t)}_j}, &\text{if} $k \in \mathcal{U}_t$\\
        0, & otherwise%
        \end{IEEEeqnarraybox}\right.
        \label{eq:prob:residual}
        \end{equation}
        \State pick $i^{(t)}\in\mathcal{U}_t$ based on $\brm{p}^{(t)}$%
        \State {Kaczmarz update step (Algo. \ref{algo:RK} -- Steps 10-14)}\,{\%{ $8M+4$ FLOPs}}%
        \EndWhile
        \State $\brm{\hat{x}}^{\text{GRK}}=\brm{v}^{(t)}$, $T^{\text{GRK}}=t$
    \end{algorithmic}
\end{algorithm}

We start by explaining the functionality of the GRK-RZF scheme. At the beginning of the iterative approach, the current residual vector $\brm{r}^{(t)}$ is calculated in Step \ref{algo:GRK:res}. We refer to $\brm{r}^{(t)}$ as the \textit{complete residual information} on an iteration basis. In Steps 5 and 6, we obtain the squared absolute residuals arranged in a vector $\vv{\mathrm{SAR}}^{(t)}$ and the residual sum of squares $\rss^{(t)}$. Then, in Step 7, the quantity $\epsilon^{(t)}$ as a measure of the weighted average of the normalized squared absolute residuals is computed. Using this quantity, in Step 8, we can select a set $\mathcal{U}_t$ of \textit{working equations} that corresponds to the equations with residuals larger than the weighted average. The idea is to discard equations with the lowest residuals prioritizing the equations (users) that are further away from being solved. In Step 9, the sampling probability vector $\brm{p}^{(t)}$ is now iteration dependent and calculated in \eqref{eq:prob:residual} on the basis of the squared absolute residuals of the equations in $\mathcal{U}_t$. The subsequent steps follow the Kaczmarz update step. When the algorithm converges, we obtain the GRK-RZF soft estimate $\brm{\hat{x}}^{\text{GRK}}$. A relaxed version of the GRK algorithm is presented in \cite{Bai2018}, where one can control the quantity $\epsilon^{(t)}$ and adjust the size of $\mathcal{U}_t$. However, we chose not to follow this method because the control of $\epsilon^{(t)}$ can generate unwanted complexity.

\subsubsection{Probability criterion based on complete residual information} The GRK-RZF exploits the \emph{complete residual information} as part of its probability criterion in \eqref{eq:prob:residual} used to select the working equations (preferred users) from the SLE in \eqref{eq:consistent_linearsys}. Evidently, this use can solve the two fundamental problems of the nRK-RZF scheme \cite{Boroujerdi2018a} of rare equations and curse of uniform normalization in a heuristic way because the residuals progress as solutions become better. The trend is that the residuals tend to zero as the number of iterations grows towards infinity. Therefore, it is expected that the number of necessary iterations for convergence of the GRK-RZF scheme is less than that of the RK-RZF. However, obtaining the complete residual information and its processing makes iteration more expensive. This indeed can lead to the case where the total complexity cost of the GRK-RZF receiver after convergence is greater than that of the RK-RZF and even of the RZF. This issue driven us to look for ways to further explore the performance gains brought by the residuals and reduce the related cost. We further elaborate the benefits brought by the probability criterion based on the residuals in Subsection \ref{subsec:interpret:residual}.

The first way to decrease the complexity of the GRK-RZF scheme is to adopt the following recursive relationship \cite{Bai2018a}:%
\begin{IEEEeqnarray}{rCl}
    \brm{r}^{(t+1)}&=&\brm{b}-\Hhat^\htransp\brm{u}^{(t+1)}-\xi\brm{v}^{(t+1)}\IEEEnonumber\\%
    &\stackrel{\text{(a)}}{=}&\brm{b}-\Hhat^\htransp(\brm{u}^{(t)}+\gamma^{(t)}\brm{h}_{i^{(t)}})-\xi(\brm{v}^{(t)}+\gamma^{(t)}[\eye{K}]_{:,i^{(t)}})\IEEEnonumber\\%
    &=&\brm{b}-\Hhat^\htransp\brm{u}^{(t)}-\xi\brm{v}^{(t)}-\gamma^{(t)}\Hhat^\htransp\brm{h}_{i^{(t)}}-\gamma^{(t)}\xi[\eye{K}]_{:,i^{(t)}}\IEEEnonumber\\%
    &\stackrel{\text{(b)}}{=}&\brm{r}^{(t)}-\gamma^{(t)}(\Hhat^\htransp\brm{h}_{i^{(t)}}+\xi[\eye{K}]_{:,i^{(t)}})\IEEEnonumber\\%
    &\stackrel{\text{(c)}}{=}&\brm{r}^{(t)}-\gamma^{(t)}[\brm{R}_{\brm{yy}}]_{:,i^{(t)}},\label{eq:recurs}%
\end{IEEEeqnarray}
where the following steps were applied: (a) the Kaczmarz iteration relationship (Algo. \ref{algo:GK} -- Step 11), (b) the definition of the residual vector $\brm{r}^{(t)}\in\mathbb{C}^{K \times 1}$ at iteration $t$ (Algo. \ref{algo:GK} -- Step 4), and (c) the definition of $\brm{R}_{\brm{yy}}$ in \eqref{eq:zf_rzf}. Henceforth, {we assume that the GRK-RZF scheme adopts the above recursive updating of the complete residual information.}

\subsection{Randomized Sampling Kaczmarz Algorithm}
To reduce the complexity related to the processing of residuals and still exploit part of this information, Algorithm \ref{algo:RSK} describes the RSK method proposed in \cite{Sun2020} to solve the SLE in \eqref{eq:consistent_linearsys}. We called this as the RSK-RZF scheme. Here, the iterative process starts by uniformly drawing equations to comprise the set $\mathcal{U}_t$ of working equations with a pre-defined size of $\omega$. Subsequently, only the residuals of these $\omega$ equations are calculated, reducing the iteration cost in comparison with the GRK-RZF scheme. To select the equation at iteration $t$, a deterministic criterion is now adopted: the equation $k\in\mathcal{U}_t$ with the largest entry in the relative residual vector $\vv{\mathrm{RR}}^{(t)}$ is chosen. Then, the algorithm follows the Kaczmarz update step. 

\subsubsection{Probability criterion based on partial residual information} 
Different from the other algorithms, randomization is used  when constructing $\mathcal{U}_t$ and sorting $\vv{\mathrm{RR}}^{(t)}$ and depends on a \textit{partial residual information} coming from the $\omega$ equations selected at random. Because of this, the RSK-RZF scheme gives up some performance gains, since $\mathcal{U}_t$ may not have the equations (users) that have the largest residuals. Another implementation issue that arises with Algorithm \ref{algo:RSK} is how to select $|\mathcal{U}_t|=\omega$. Motivated by \cite{Sun2020}, we use $\omega=\lceil\log_2 K\rceil$.

\begin{algorithm}[t] 
    \caption{RSK-Based Receiver (RSK-RZF)\label{algo:RSK}}
    \algrenewcommand\algorithmicrequire{\textbf{input:}}
    \algrenewcommand\algorithmicensure{\textbf{output:}}
    \begin{algorithmic}[1]
        \Require$\brm{H}$, $\brm{y}$, $M$, $K$, $\xi$, $\omega$%
        \Ensure$\brm{\hat{x}}^{\text{RSK}}$, $T^{\text{RSK}}$%
        \State Repeat Steps 1-6 of Algorithm \ref{algo:RK}{\%{ $16KM-2K-1$ FLOPs}}%
        \State Store $(\fronorm{\brm{H}}^{2}+K\xi)^{-1}$\qquad{\%{ $1$ flop}}%
        \While{$\text{stopping criterion}\text{\textbf{ is }}\mathrm{False}$}\label{algo:RSK:termination}%
        \State Uniformly draw w/ SwoR $\mathcal{U}_t$ w/ $|\mathcal{U}_t|=\omega$\label{algo:RSK:wor}%
        \State $\brm{r}^{(t)}\in\mathbb{C}^{K\times{1}}$ with\qquad{\% $\omega(8M+4)$ FLOPs}\label{algo:RSK:res}
        \begin{equation}
        \setlength{\nulldelimiterspace}{0pt}
        r^{(t)}_{j}=\left\{\begin{IEEEeqnarraybox}[\relax][c]{l's}
        b_j-\brm{h}_{j}^\htransp \brm{u}^{(t)} - \xi v^{(t)}_j, &\text{if} $j\in\mathcal{U}_t$\\
        0, & otherwise%
        \end{IEEEeqnarraybox}\right.\nonumber
        \end{equation}
        \State Compute $\vv{\mathrm{RR}}^{(t)}\in\mathbb{R}^{K\times 1}$ w/\qquad\%{ $4\omega$ FLOPs}%
        $$\mathrm{RR}^{(t)}_{k}=|r^{(t)}_k|^2/(\fronorm{\brm{H}}^{2}+K\xi)\,\forall{k}\in\mathcal{K}$$%
        \State $i^{(t)}=\argmax_{j\in\mathcal{U}_t}\vv{\mathrm{RR}}^{(t)}$\quad\,\,\,\,\%{ $\omega$ FLOPs}
        \State {Kaczmarz update step (Algo. \ref{algo:RK} -- Steps 10-14)}\,{\%{ $8M+4$ FLOPs}}%
        \EndWhile%
        \State $\brm{\hat{x}}^{\text{RSK}}=\brm{v}^{(t)}$, $T^{\text{RSK}}=t$
    \end{algorithmic}
\end{algorithm}

\section{RK-Based Receiver Designs for Multi-Antenna Systems}\label{sec:interpret}
This section addresses two issues on the algorithms' operation described above when applied to the M-MIMO context: What is the meaning of a RK-based iteration for multi-antenna systems? What are the main benefits of using residual information in the equation selection probability criterion? A short answer to both of the questions is: we notice a similar operation of our schemes with the SIC and we find robustness against IUI and sparsity, respectively. We elaborate further on each of these issues in the sequel.

\begin{figure}[!htbp]
    \centering
    \includegraphics[trim=0.7cm 0.1cm 0.7cm 0cm, clip=true, width=\linewidth]{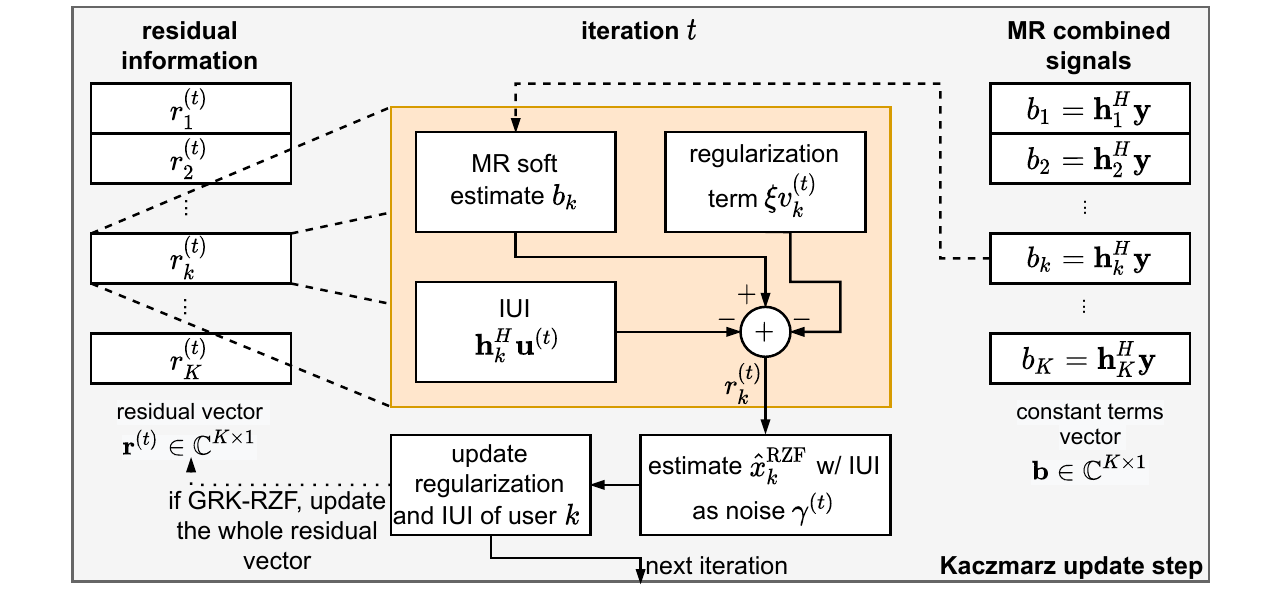}
    \vspace{-6mm}
    \caption{Illustration of the Kaczmarz update step when user $k$ is selected at iteration $t$. Observe that the residual $r^{(t)}_k$ stores the MR combined signal of user $k$ less: (i) IUI from previous iterations and (ii) the regularization term applied to combat noise.}
    \label{fig:residual-progress}
\end{figure}

\begin{table*}[htp]
    \centering
    \caption{{Computational Complexity Comparison}}
    \label{tab:cc}
    \resizebox{1\textwidth}{!}{%
    \begin{tabular}{|c|c|c|c|}
        \hline
        \textbf{Scheme} & \textbf{Computational Complexity [FLOPs]} &  \begin{tabular}[c]{@{}c@{}} \textbf{M-MIMO [FLOPs]} \\($M=64,K=8,T=12$)$^{\dagger}$\end{tabular}  & \begin{tabular}[c]{@{}c@{}} \textbf{XL-MIMO [FLOPs]} \\($M=256,K=32,T=64$)$^{\dagger}$\end{tabular} \\ \hline
        MR & $8KM-2K$ & $4080$ & $65472$\\ \hline 
        RZF & $4K^2M+12KM+5K^3+10K^2-4K$ & $25696$ & $1320832$ \\ \hline
        nRK-RZF \cite{Boroujerdi2018a} & $16KM-K-1+(16M+8){T^{\text{nRK}}}$ & $20567$ & $393695$ \\ \hline
        \begin{tabular}[c]{@{}c@{}}RK-RZF \\(Algorithm \ref{algo:RK})\end{tabular} & $16KM-2K-1+(K+16M+8){T^{\text{RK}}}$ & $20653$ & $395711$\\ \hline
        \begin{tabular}[c]{@{}c@{}}GRK-RZF\\(Algorithm \ref{algo:GK})\end{tabular} & $4K^2M+12KM-K^2-K+(16K+8M+7){T^\text{GRK}}$ & $30220$ & $1310112$\\ \hline
        \begin{tabular}[c]{@{}c@{}}RSK-RZF\\(Algorithm \ref{algo:RSK})\end{tabular} & $16KM-2K+[\omega(8M+9)+8M+4]{T^\text{RSK}}$ & $33124$ & $920576$ \\ \hline
        TPE-RZF \cite{Sessler2005,Kammoun2014} & $4K^2M+12KM+3K+4+(8K^2+4K){T^{\text{TPE}}}$ & $29696$ & $1679460$ \\ \hline
    \end{tabular}%
    }
    \flushleft {\footnotesize $^\dagger T=T^{\text{nRK}}=T^{\text{RSK}}=T^{\text{GRK}}=T^{\text{RSK}}=T^{\text{TPE}}$; $T^s$ denotes the number of iterations of scheme $s\in\{\text{nRK},\text{RSK},\text{GRK},\text{RSK},\text{TPE}\}$.}%
\end{table*}

\subsection{A Similarity with SIC receiver}\label{subsec:interpret:sic}
We show that the structure of the RK algorithms applied to solve the SLE in \eqref{eq:consistent_linearsys} yields a mechanism similar to the used by the SIC receiver. For that, Fig. \ref{fig:residual-progress} illustrates in details the Kaczmarz update step that is common to all algorithms. In this figure, we assume that user $k\in\mathcal{K}$ is selected at a given iteration $t$. In addition, the residual vector $\brm{r}^{(t)}$ and the vector of constant terms $\brm{b}$ with the MR soft estimate are represented by stacks of blocks. An interesting pattern can be observed from the figure. The residual $r^{(t)}_k$ is storing the MR combined signal of user $k$ less: (i) the IUI from previous iterations and (ii) the regularization term applied to combat noise. Then, $r^{(t)}_k$ is used to get a new estimate of $\hat{x}^{\text{RZF}}_k$ considering remaining IUI as noise. We notice that this mechanism is similar to that used by the SIC receiver \cite{Carvalho2012}, which successively remove the contribution of the decoded data from the received signal on an iteration basis. Mathematically, we can obtain this interpretation of the residual using the recursive relationship in \eqref{eq:recurs}. From this, we can write
\begin{IEEEeqnarray}{rCl}
r^{(t)}_k&=&\brm{h}_k^\htransp\brm{y}-\sum_{t'=1}^{t}\gamma^{(t')}[\brm{R}_{\brm{yy}}]_{k,i^{(t')}}  \IEEEnonumber\\%
&=&\brm{h}_k^\htransp\brm{y}-\sum_{t'=1}^{t}\gamma^{(t')}\{\underbrace{\chi_{\mathcal{I}}(i^{(t')})\mathbf{h}^\htransp_k\mathbf{h}_{i^{(t')}}}_{\text{IUI}} + \IEEEnonumber \\
&+&
[1-\chi_{\mathcal{I}}(i^{(t')})]\underbrace{[\ltwonorm{\brm{h}_k}^2 + \xi]}_{\text{self-knowl. + reg.}}\}\label{eq:residual-relationship}
\end{IEEEeqnarray}
where $\chi_\mathcal{I}(i^{(t')})$ is the indicator function with argument denoting the equation selected at iteration $t'$ and $\mathcal{I}=\mathcal{K}\setminus\{k\}$ is the set of interfering users in relation to user $k$. In the expression above, we used the fact that $r^{(0)}_k=b_k$ since $\brm{u}^{(0)}$ and $\brm{v}^{(0)}$ are initialized with zeros. Moreover, it can be seen that: if user $k$ was selected at previous iterations, the previous estimates of $\hat{x}^{\text{RZF}}_k$ is also removed from $r_k^{(t)}$ together with the noise penalization (self-knowledge + regularization in \eqref{eq:residual-relationship}). Recall that the RK-based algorithms are approximating the RZF scheme when solving the SLE in \eqref{eq:consistent_linearsys}. The application of the RK algorithms over \eqref{eq:consistent_linearsys} is then transforming the RZF scheme into a SIC-alike receiver. Effectively, we are refining the MR soft estimate $\hat{\brm{x}}^{\text{MR}}$ stored in $\brm{b}$ to approach the RZF soft estimate $\hat{\brm{x}}^{\text{RZF}}$, placing the SIC-alike iterations in charge of computing the weights of IUI suppression based on the RZF criterion. This adaptive mechanism implicitly helps supporting more users in multi-antenna systems with low-complexity. In addition, the RK-based receivers do no explicitly calculate metrics, such as post-processing SNR or signal-to-interference-plus-noise ratio (SINR), that are normally used to order the classical SIC receiver \cite{Carvalho2012}. In fact, the RK-based algorithms give preference to the equations (users) based on the two types of information promptly provided by the SLE in eq. \eqref{eq:consistent_linearsys}: the energy information of the equations in \eqref{eq:prob:energy} and the complete residual information in \eqref{eq:prob:residual}, which is partial for the RSK-RZF. We discuss the advantages of the latter below. 

\subsection{Probability Criterion and Residual Information}\label{subsec:interpret:residual}
The probability criterion in \eqref{eq:prob:residual} uses the complete residual information $\brm{r}^{(t)}$ to select the equations of \eqref{eq:consistent_linearsys} to be solved. The residual contains inner products between channel vectors; hence, eq. \eqref{eq:residual-relationship} describes the IUI and naturally introduces the sparse structure arising from the spatial non-stationarities. This is a relevant contrast compared to the probability criterion in \eqref{eq:prob:energy} that has probabilities proportional to $\ltwonorm{\brm{h}_k}^2 + \xi, \forall k \in \mathcal{K}$ only. As a result, the probability criterion of \eqref{eq:prob:residual} can better capture the interaction effects of IUI and sparsity existing in the XL-MIMO channels, outperforming \eqref{eq:prob:energy} under the occurrence of theses effects. Also, \eqref{eq:prob:residual} is dynamic in the sense that the IUI terms and consequently the probabilities are updated in the background according to the SIC-alike IUI suppression mechanism running in the foreground, changing the set $\mathcal{U}_t$ of preferred users as the solution evolves. {The RSK-RZF scheme partly features these gains.} However, a clear issue of using residual information is the cost of such more involved probability criterion, as best evidenced in the next sections.

\section{Complexity Analysis and Numerical Results}\label{sec:cc-num-res}
In this section, we first discuss the computational complexity of the proposed RK-based RZF receivers in terms of floating-point operations per second (FLOPs) and how their stopping criterion can be defined in practice. The performance of the proposed receivers is numerically evaluated in the sequel by taking the bit-error-rate (BER) as a metric. Further, we assume that $\brm{x}$ is drawn from the equiprobable 16-QAM constellation. Besides MR, RZF, and nRK-RZF \cite{Boroujerdi2018a}, in the numerical simulations we compare our proposed schemes to the TPE-RZF receiver of \cite{Sessler2005,Kammoun2014}. The choice for the TPE-RZF is due to the fact that this receiver is a consolidated approach in the literature that also iteratively approximates the RZF scheme. Finally, for tractability reasons, we further consider horizontal uniform linear array (ULA) arrangements under non-line-of-sight conditions in both M-MIMO and XL-MIMO regimes, {with the distance between any two neighboring antennas greater than half a wavelength when considering sub-6 GHz transmissions}.

\subsection{Complexity Analysis}\label{subsec:cc}
The second column of Table \ref{tab:cc} summarizes the total number of FLOPs needed to compute the receiver designs relevant for this work. We always account for the worst-case when performing the complexity analysis of our proposed schemes. For Algorithm \ref{algo:RK}, this means that the cost of Step \ref{algo:RK:swor} is considered to be $K$ FLOPs at most due to the re-normalization of $\brm{p}$. For Algorithm \ref{algo:GK}, Step \ref{algo:GRK:probability} costs $K$ FLOPs, since $|\mathcal{U}_t|$ is always considered to be $K$. Complexity of the MR and RZF schemes follows \cite{Bjornson2017c}, while the complexity of the TPE-RZF receiver is discussed in detail in \cite{Sessler2005} and \cite{Kammoun2014}. For the TPE-RZF, we adopt the eigenvalue estimation of $\brm{R}_{\brm{yy}}$ proposed in \cite{Sessler2005}. For the sake of fairness, we assume that the BS does not know second-order channel statistics. Thus, the scaling proposed in \cite{Sessler2005} to lower the scattering of the eigenvalue is not performed. This allows us to show that our methods are more robust to channel gain variations between users without the need to seek alternatives to reduce the impact of these variations.

To give a better notion of how the computational complexities are compared to each other, we evaluate two typical scenarios of both M-MIMO and XL-MIMO regimes in Table \ref{tab:cc}. Note that we set the same number of iterations $T$ for all the iterative algorithms in the table, where $T=12$ and $T=64$ iterations for M-MIMO and XL-MIMO, respectively. We chose these numbers because they allow us to show the computational gains brought by our receiver designs, while achieving good performance. We notice the hereafter trends from the table{: {\bf a})} for M-MIMO, the GRK-RZF scheme is unable to relax the RZF, exemplifying the high cost of complete residual information{; {\bf b})} in contrast, the RK-RZF receiver can relax complexity of the RZF in 19.62\%. For XL-MIMO{: {\bf c})} the GRK-RZF scheme is up to reduce the complexity of the RZF in 0.81\%, while the RK-RZF achieves a relaxation of 70.19\%; {\bf d}) the RSK-RZF scheme only achieves its goal of relaxing the GRK-RZF in the XL-MIMO scenario; {\bf e}) the TPE-RZF receiver has iterations independent of $M$, but it has a high fixed cost due to the exact computation of $\brm{R}_{\brm{yy}}$ and the estimation of its eigenvalues. From these observations, we noticed that as $M$ and $K$ increase, the relaxation capacity of the GRK-RZF receiver is improved. However, the RK-RZF will always have a greater ability to relax complexity, since its iterations are cheap. 

Next, we evaluate the difference in performance among the proposed receivers, identifying when the use of residual information becomes justified.

\subsubsection{Sparsity} In the case of a sparse SLE in \eqref{eq:consistent_linearsys}, we can automatically reduce the costs of the iterations of the RK-based receivers. The reason for this is to note that inner products are the most cost operations in the iterations of all the algorithms, which can be evidently reduced by only using the non-zero entries of the vectors.

\subsubsection{Defining in practice the number of iterations}
The most convenient stopping criterion of all the proposed algorithms is the maximum number of iterations. The BS can regularly adjust the maximum number of iterations after a constant period of time-frequency resources that spans multiple coherence blocks. This adjustment can be based on some performance or complexity metric that the BS wants to achieve.

\subsection{Stationary Case: M-MIMO}\label{subsec:stationary}
Consider a cell that covers a square area of $0.4$ km $\times$ $0.4$ km served by a BS with $M=64$ compactly installed antennas {located at the cell center}. The users are uniformly distributed in the cell area at locations further than 35 m from the BS. Furthermore, for this scenario we assume that all the elements of $\boldsymbol{\beta}_k$ are equal, since the distance between antennas is much smaller than the distance between users and the antenna array. Then, we model the pathloss based on the urban micro scenario as \cite{Bjornson2017c}: $\beta_k=-30.5-36.7\log_{10}d_k$ in dB, where $\beta_k=\frac{1}{M}\trace{\brm{R}_k}$ is the average large-scale coefficient, $d_k$ is the distance in meters between user $k\in\mathcal{K}$ and the BS. In addition, we consider the more general exponential correlation model in which $[\brm{R}_k]_{i,j}=\iota^{\lvert i-j \rvert}, \ \forall i,j\in\mathcal{M}$ \cite{Bjornson2018}, where $\iota$ is the antenna correlation coefficient.

\begin{figure}[htp]
    \centering
    \subfloat[Average BER versus iterations.\label{fig:mmimo:conv:performance}]{\includegraphics[trim=4mm 3mm 3mm 3mm,clip,width=\linewidth]{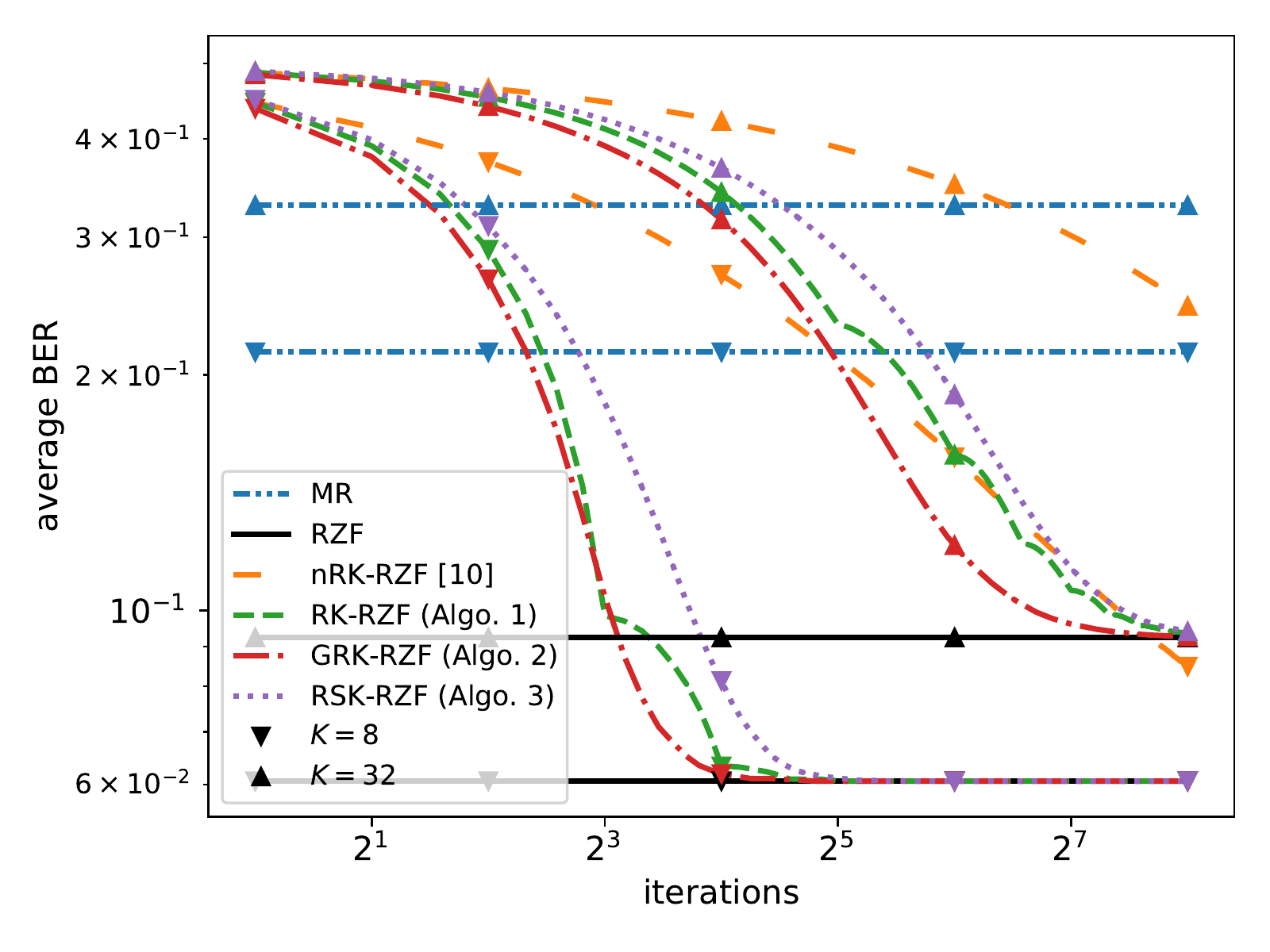}}\quad
    \subfloat[Computational complexity in FLOPs versus iterations.\label{fig:mmimo:conv:complexity}]{\includegraphics[trim=4mm 3mm 3mm 3mm,clip,width=\linewidth]{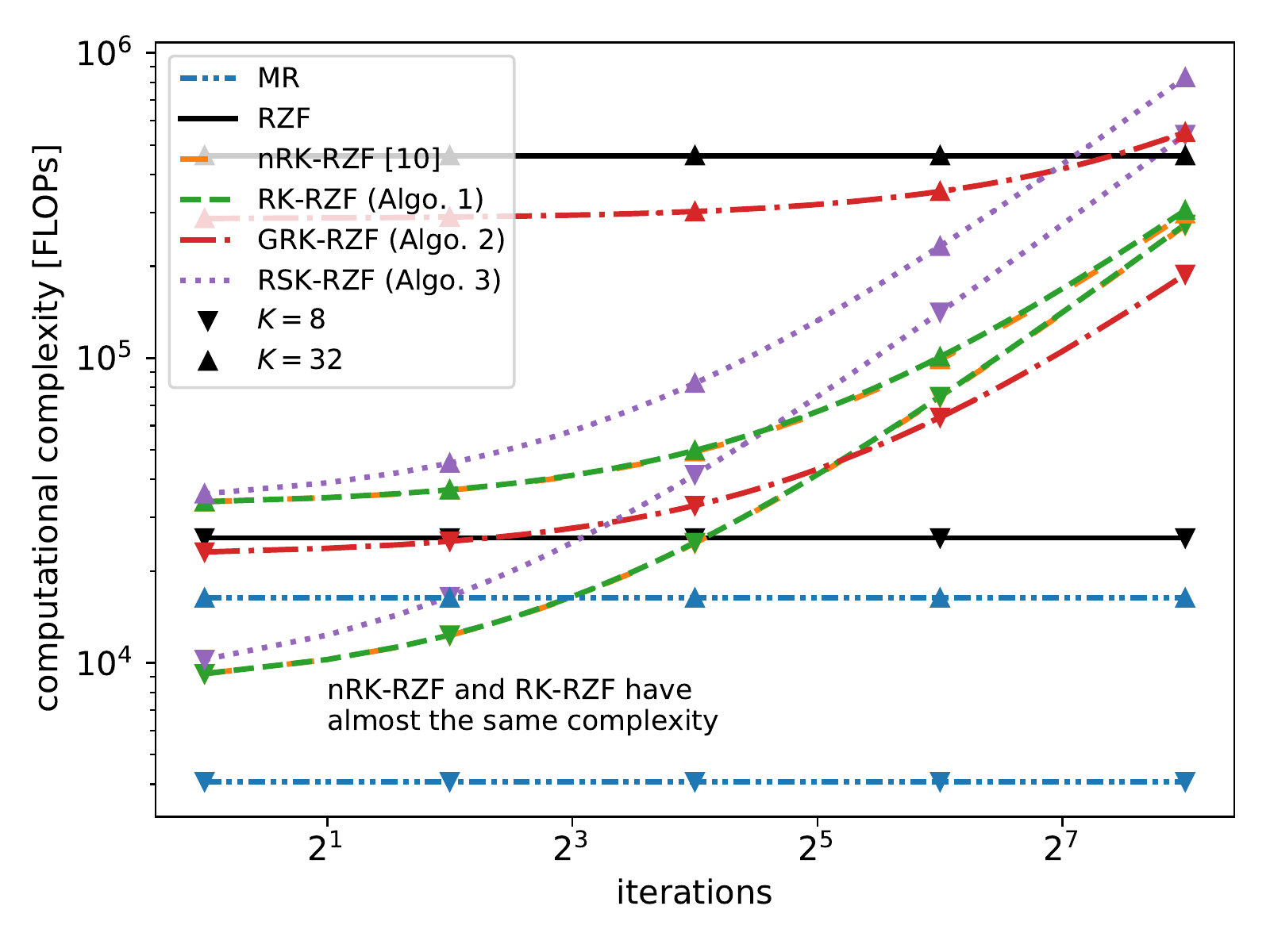}}    
    \caption{Convergence of the 
    RK-based RZF schemes in a normally loaded ($K=8$) and crowded ($K=32$) M-MIMO system with $M=64$ antennas, under Rayleigh fading channels, and a pre-processing SNR of 0 dB.}
    \label{fig:mmimo:conv}
\end{figure}

We first evaluate the convergence in terms of both performance (Fig. \ref{fig:mmimo:conv:performance}) and complexity (Fig. \ref{fig:mmimo:conv:complexity}) of the proposed algorithms. A typical load of $K=8$ users and a crowded scenario with $K=32$ users are examined. The first observation from Fig. \ref{fig:mmimo:conv:performance} is that the performance of our three accelerated RK-based RZF schemes is much better than that obtained with the nRK-RZF proposed in \cite{Boroujerdi2018a}. We also notice that increasing the IUI with increasing $K$ harms the convergence of RK-based schemes in general. However, the GRK-RZF suffers the least from increased IUI. This result is inline with the observed fact that randomization based on the residual information is more robust against IUI. However, we learn from Fig. \ref{fig:mmimo:conv:complexity} that the GRK-RZF receiver is not suitable for typical M-MIMO scenarios in terms of complexity and only starts to get more appealing in this regime as $K$ increases. {Moreover, the bouncy behavior of the performance curves associated to RK-RZF in Fig. \ref{fig:mmimo:conv:performance} is explained by the SwoR technique and the stochastic behavior of the elements of the set $\mathcal{P}^{(t)}$. Note that the start and end point of the bounce comprehends the definition of a sweep made in Subsection \ref{subsec:rk:algo} that embraces $K$ iterations.}

\begin{figure}[htp]
    \centering
    \includegraphics[trim=4mm 3mm 3mm 3mm,clip, width=\linewidth]{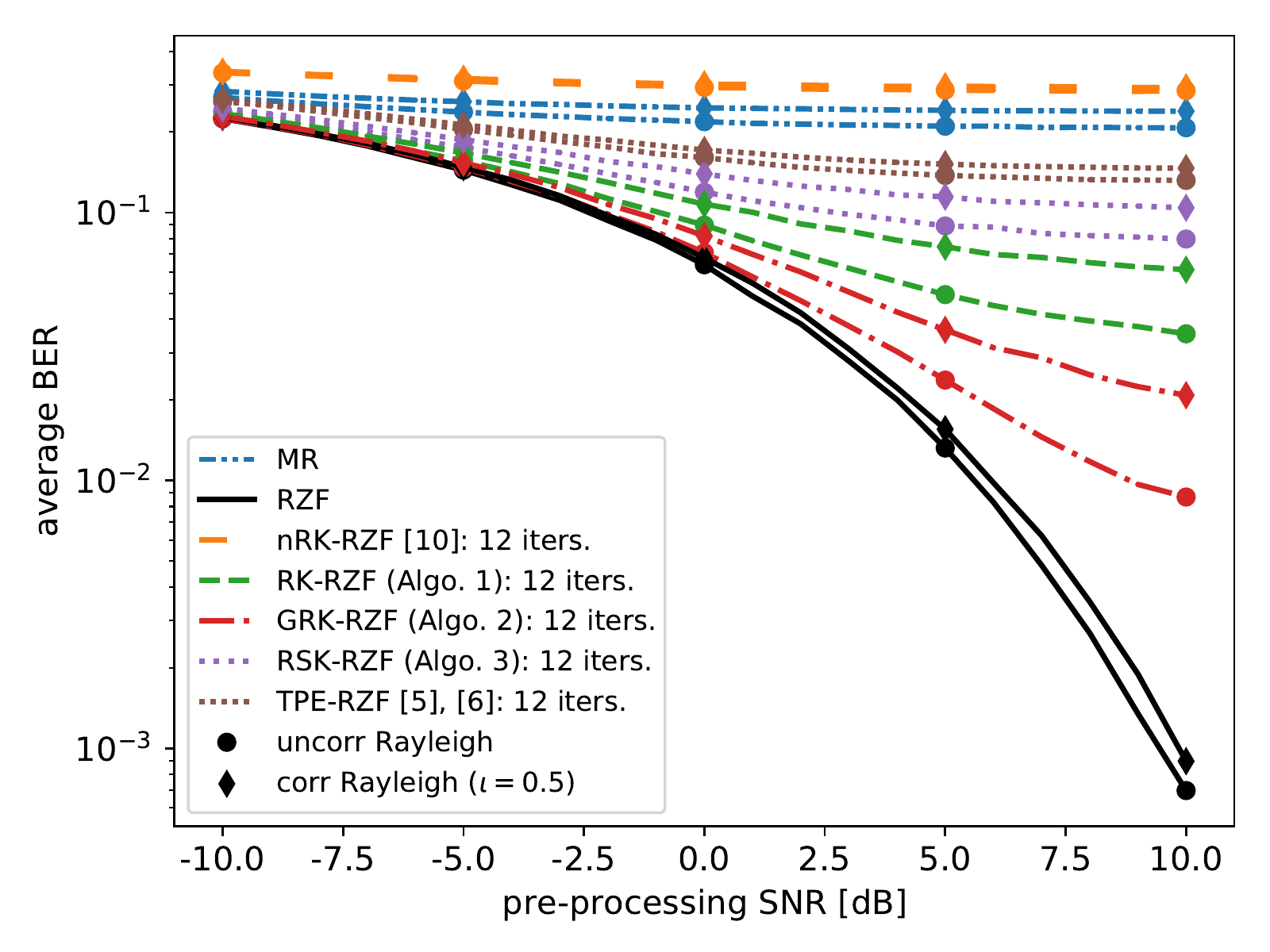}
    \vspace{-6mm}
    \caption{Average BER performance of various receivers {\it vs.} the pre-processing SNR under uncorrelated and correlated ($\iota=0.5$) Rayleigh channels for the M-MIMO system equipped with $M=64$ antennas and $K=8$ users.}
    \label{fig:mmimo:snr}
\end{figure}

Fig. \ref{fig:mmimo:snr} depicts the BER performance of different receivers as a function of the pre-processing SNR $\rho/\sigma^2$ with $K=8$ users for uncorrelated and correlated ($\iota=0.5$) Rayleigh fading conditions. It considers a number of iterations fixed in 12 for all the iterative schemes. The final complexity of each scheme follows the third column in Table \ref{tab:cc}. Among the RK-based RZF schemes, the GRK-RZF attains the best performance, but needs more FLOPs than the RZF scheme. In contrast, the RK-RZF performs well as a whole while relaxing the complexity of the RZF in 19.62\%. In general, the accelerated RK-based RZF schemes better approximate the performance of the RZF at low pre-processing SNRs. This is because the strength of IUI is amplified when operating at high SNRs. Finally, our schemes perform better than the TPE-RZF scheme \cite{Sessler2005,Kammoun2014} with less complexity.

\subsection{Non-Stationary Case: XL-MIMO}\label{subsec:non-stationary}
Let's consider a square cell with an area of $0.25$ km $\times$ $0.25$ km that has totally occupying one of its side by a ULA equipped with $M=256$ antennas. The users are uniformly distributed in the cell keeping a minimum distance of $25$ m from the array.\footnote{{The choice for these values and the geometry is motivated by the fact that the users are close enough to the array to justify the emergence of spatial non-stationarities \cite{Carvalho2012,Amiri2020}. It is noteworthy that the adopted geometry of M-MIMO and XL-MIMO are comparable given that in M-MIMO, the BS is cell-centered; while in XL- MIMO geometry, the BS comprises one of the edges of the square area.}} The distance between the user and antenna elements is now relevant. Therefore, each element $\beta^{m}_k$ of $\boldsymbol{\beta}_k$ is modeled as $\beta^{m}_k=-30.5-36.7\log_{10}d^{m}_k$, where $d^{m}_k$ is the distance between user $k\in\mathcal{K}$ and antenna $m\in\mathcal{M}$. Under this setting, we focus on non-stationarities and consider antenna correlation irrelevant, then $\brm{R}_k=\eye{M}$. Moreover, we generate $\brm{D}_k$ for user $k\in\mathcal{K}$ as follows \cite{Ali2019a}: {\bf a}) we choose an arbitrary antenna $m\in\mathcal{M}$ uniformly at random to be the center $c_k$ of the VR{; \bf b}) if $D$ is odd, the VR of user $k$ is $\mathcal{V}_k=\{c_k-\lfloor D/2\rfloor, \dots, c_k+\lfloor D/2\rfloor\}$, otherwise $\mathcal{V}_k=\{c_k-\lfloor D/2\rfloor, \dots, c_k+\lfloor D/2\rfloor+1\}${; \bf c}) we set $[\brm{D}_k]_{m,m}=1$, if $m\in\mathcal{V}_k\cap\mathcal{M}$ and $[\brm{D}_k]_{m,m}=0$ otherwise, and {\bf d}) we normalize $\brm{D}_k$ by $M/D$, hence stationary and non-stationary channels have the same norm \cite{Ali2019a}. We stress that this normalization is giving an array gain for non-stationary channels similar to the stationary channels, allowing a fair comparison between the two array regimes. With this model, the users have a unique cluster of antennas representing their VRs with an average size of $D$. To evaluate how the receivers behave under very extreme sparse conditions, in the following, we set low values for $D$. In \cite{Ali2019a}, for example, the authors report problems with the ZF scheme from $D$ below $30$ visible antennas. Here, the RZF regularization term makes it possible to perform the matrix inversion even in these severe conditions.

\begin{figure}[htp]
    \centering
    \subfloat[Average BER versus iterations.\label{fig:xlmimo:conv:performance}]{\includegraphics[width=\linewidth]{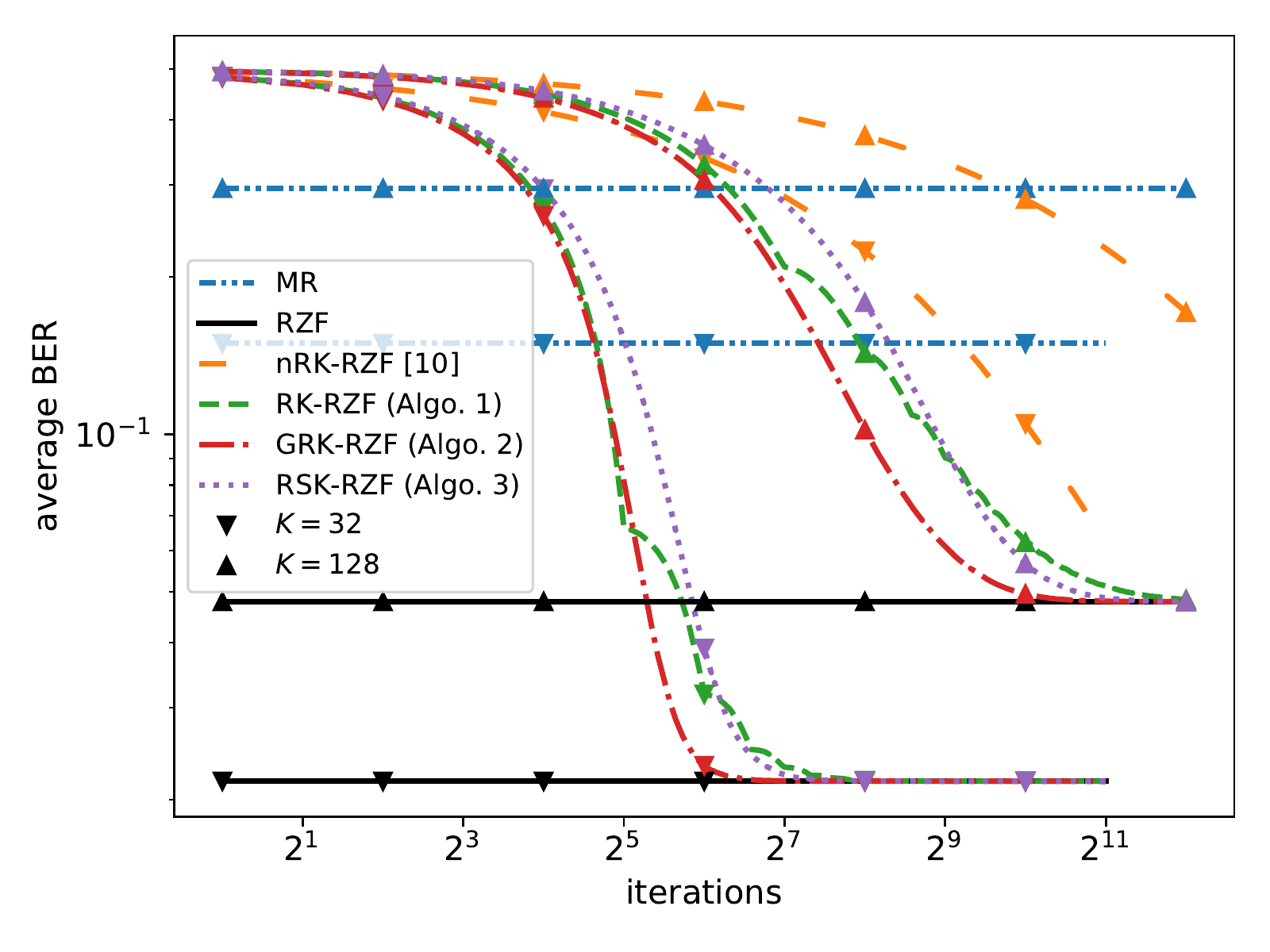}}\quad
    \subfloat[Computational complexity in FLOPs versus iterations.\label{fig:xlmimo:conv:complexity}]{\includegraphics[width=\linewidth]{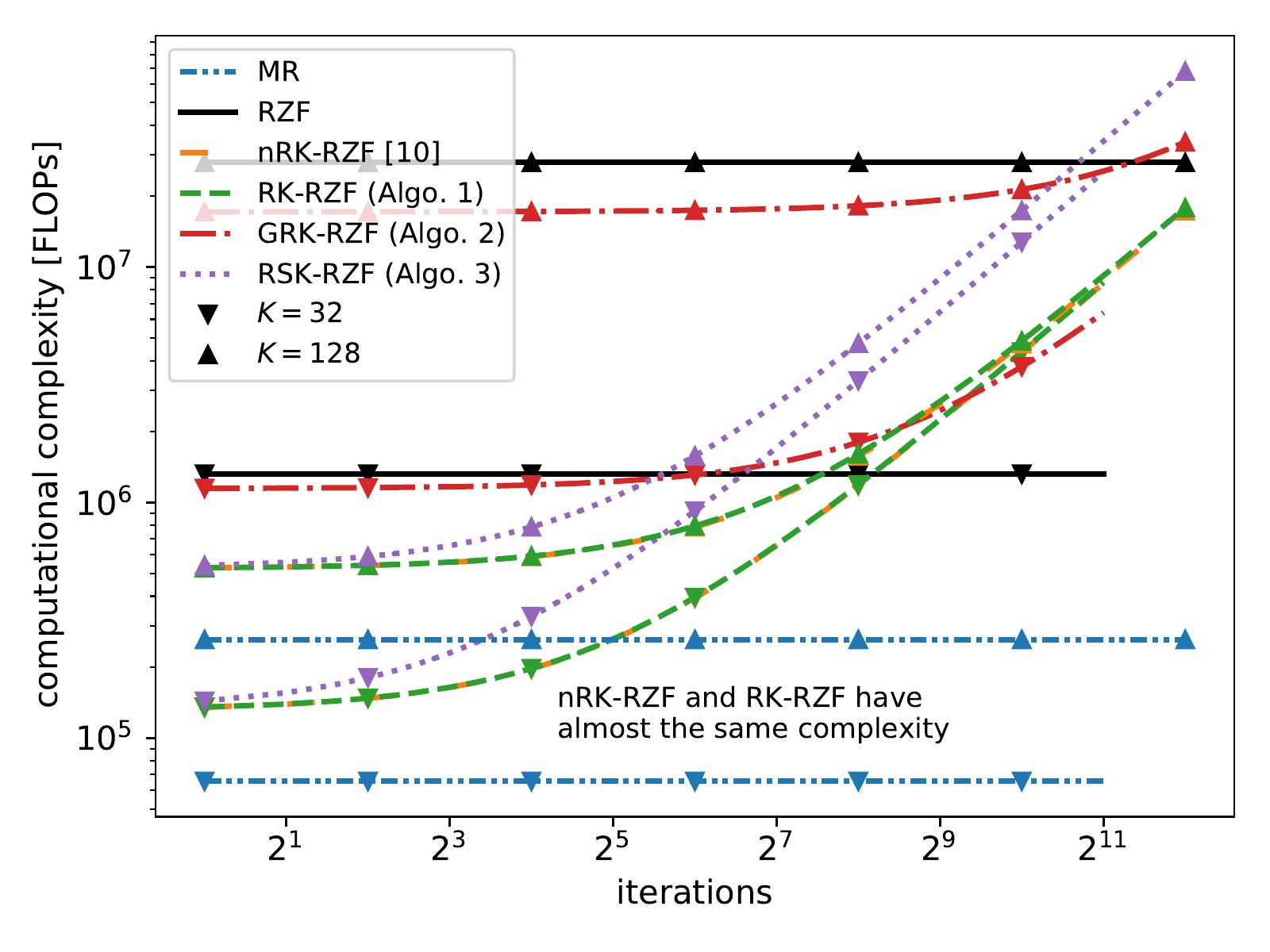}} 
    \caption{Convergence of the different RK-based RZF schemes in a normally loaded ($K=32$) and crowded ($K=128$) XL-MIMO system with $M=256$ antennas, under Rayleigh fading channels, a pre-processing SNR of 0 dB, and a sparsity level of $D=8$ visible antennas.}
    \label{fig:xlmimo:conv}
\end{figure}

We first take a look at Fig. \ref{fig:xlmimo:conv} that exhibits the convergence of the RK-based RZF schemes and how it impacts performance (Fig. \ref{fig:xlmimo:conv:performance}) and complexity (Fig. \ref{fig:xlmimo:conv:complexity}) with $D=8$. Again, a typical load of $K=32$ users and a crowded scenario with $K=128$ users are evaluated. Comparing Fig. \ref{fig:mmimo:conv:performance} and Fig. \ref{fig:xlmimo:conv:performance}, one can see the bad effects of IUI over the convergence of the algorithms and how it impacts more severely the RK-RZF scheme. Furthermore, we now find from Fig. \ref{fig:xlmimo:conv:complexity} that the GRK-RZF scheme has more room to be able to relax the RZF scheme. One of the reasons is that the values of $M$ and $K$ become higher, justifying the cost related to residual information. Another is that the RK-RZF scheme needs more iterations to achieve a better performance under high levels of IUI and sparsity.

\begin{figure}[htp]
    \centering
    \includegraphics[trim=4mm 3mm 3mm 4mm,clip, width =\linewidth]{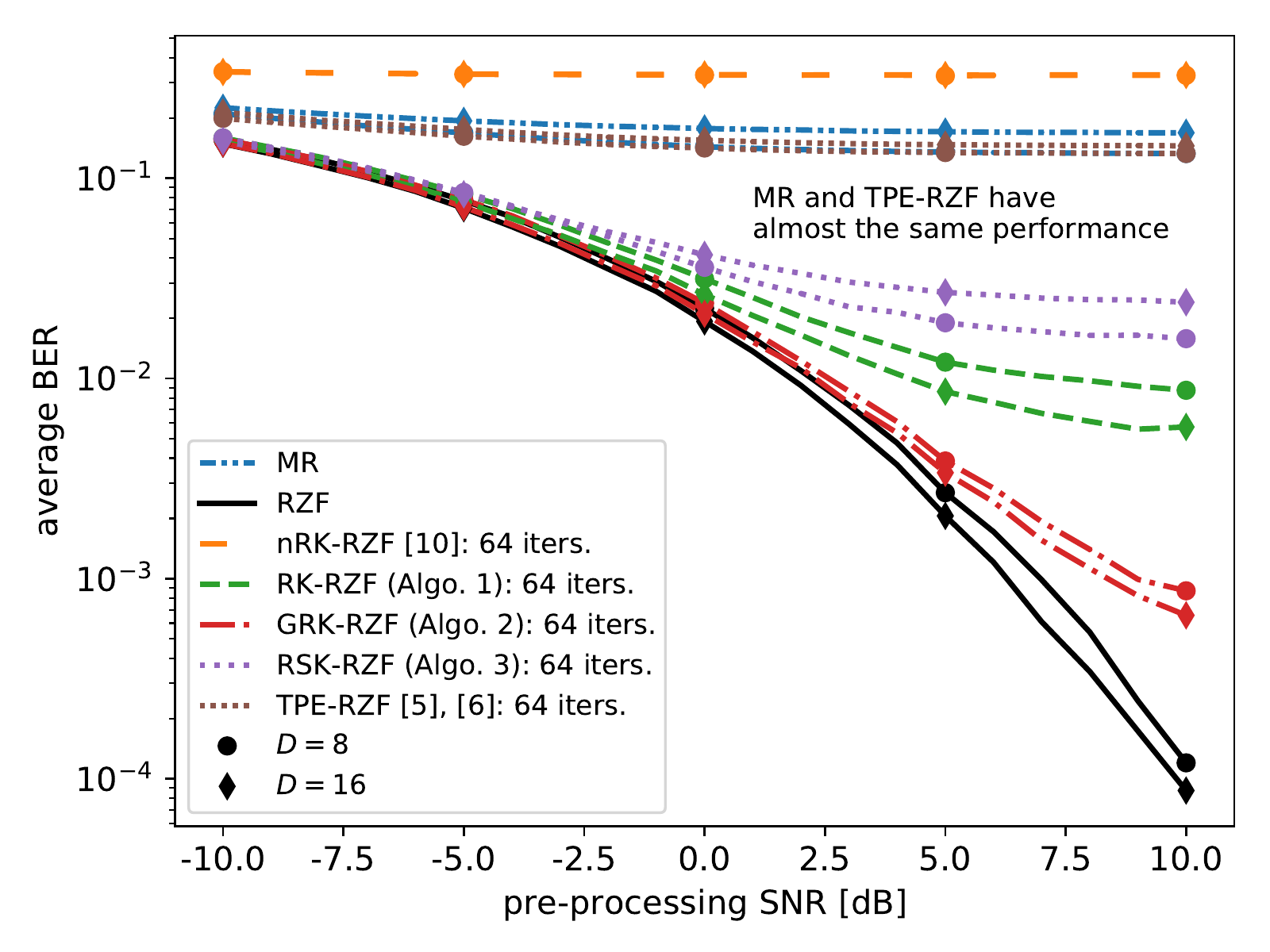}
    \caption{Average BER performance of various receivers {\it vs.} the pre-processing SNR under uncorrelated Rayleigh channels for the XL-MIMO system equipped with $M=256$ antennas and $K=32$ users; two distinct sparsity levels, $D=8$, and $D=16$ visible antennas.}
    \label{fig:xlmimo:snr}
\end{figure}

Similar to Fig. \ref{fig:mmimo:snr}, a performance comparison of the receivers is available in Fig. \ref{fig:xlmimo:snr} with $K=32$ users and under $D=8$ and $D=16$ visible antennas. In addition, we fix the number of iterations of all the iterative schemes to 64. The final complexities of every scheme is reported in the fourth column of Table \ref{tab:cc}. Definitely, we note that the performance difference between the RK-RZF and the GRK-RZF schemes increases in Fig. \ref{fig:xlmimo:snr} in comparison to Fig. \ref{fig:mmimo:snr}. This indicates that besides being more robust against IUI, the GRK-RZF scheme is more robust against the sparse structure arising from the spatial non-stationarities. Interestingly, the GRK-RZF receiver performs better than the RK-RZF in high SNR regime, while relax the RZF in 0.81\%; besides, the RK-RZF gives up of performance at high SNR and achieve a relaxation of 70.19\%. The TPE-RZF scheme \cite{Sessler2005,Kammoun2014} notably experiences a strong performance degradation from the sparse channels, corroborating a greater robustness of our proposed methods.

From the above results, we can observe that the GRK-RZF scheme becomes more convenient when the scenario is more crowded ($\uparrow K$), sparsity effects are more intense ($\downarrow$ D), and/or the system is operating at high SNR. On the other hand, the RK-RZF scheme is the most appropriate choice when more relaxation is desired, the performance losses are tolerable at high SNR, and/or the system is operating at low SNRs. {Although GRK-RZF operates better at high SNR, its gains in complexity compared to the RZF scheme may be only \textit{marginal} depending on the values of $M$ and $K$; when this is the case, the use of RZF may then be more advisable. The RSK-RZF receiver can achieve its goal of relaxing the GRK-RZF in some regions, at the cost of reduced performance; \textit{e.g.}, in the range of $2^7-2^9$ iterations for $K=128$ in Fig. \ref{fig:xlmimo:conv}.} 

\section{Conclusions}\label{sec:conclusion}
We introduced three accelerated RK-based receivers, which approximate the performance of the RZF scheme, while relaxing its complexity. In our experiments, all of our proposed schemes are able to dramatically overcome the nRK-RZF introduced in \cite{Boroujerdi2018a}. {The main feature of each scheme is in order.} The RK-RZF (Algo. \ref{algo:RK}) is the proposed receiver with the best benefit-cost ratio, performing well in typical M-MIMO and XL-MIMO circumstances, while relaxing the complexity of the RZF in almost $20\%$ and $70\%$, respectively. Moreover, the GRK-RZF scheme (Algo. \ref{algo:GK}) is more suitable for extreme cases where IUI and sparsity effects cannot be neglected. {The RSK-RZF receiver (Algo. \ref{algo:RSK}) can be more efficient than GRK-RZF in some scenarios but suffers from performance losses.}  Future work can go deeper into the theoretical analysis of the introduced receivers by using the analogy with SIC receivers revealed herein. 





\bibliographystyle{IEEEtran}
\input{main.bbl}

\vspace{-10mm}
\begin{IEEEbiography}[{\includegraphics[width=1in,height=1.25in,clip,keepaspectratio]{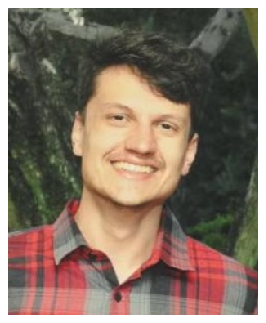}}]{Victor Croisfelt}
received his B.S. degree in electrical engineering from Universidade Estadual de Londrina (UEL), Londrina, Brazil in 2018. He is currently pursuing his M.S. degree with the Laboratory of Communications \& Signal Processing of the Department of Electrical Engineering at Universidade de São Paulo (USP), Escola Politécnica (EP), São Paulo, Brazil. He has experience in research in the areas of embedded systems and wireless communications. His current research interests are in the premises behind 6G communication systems and their optimization in different aspects.
\end{IEEEbiography}

\vspace{-10mm}

\begin{IEEEbiography}[{\includegraphics[width=1in,height=1.25in,clip,keepaspectratio]{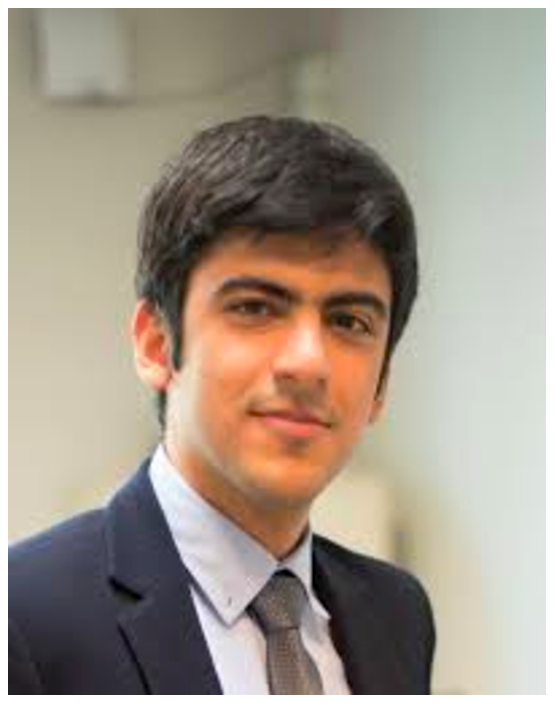}}]{Abolfazl Amiri} received his B.S. degree in Electrical and Communications Engineering (with honors) from University of Tabriz, Tabriz, Iran in 2013, and M.S. degree in Electrical and Communication Systems Engineering, from University of Tehran, Tehran, Iran in 2016. He is currently a PhD fellow in Electronic systems department of Aalborg University, Aalborg, Denmark. He was with Huawei technologies Tehran office from 2016 to 2017 as an RF engineer and cellular network optimizer. His research interests include applications of signal processing and machine learning in wireless communications and multi-antenna systems.
\end{IEEEbiography}

\vspace{-10mm}

\begin{IEEEbiography}[{\includegraphics[width=1in,height=1.25in,clip,keepaspectratio]{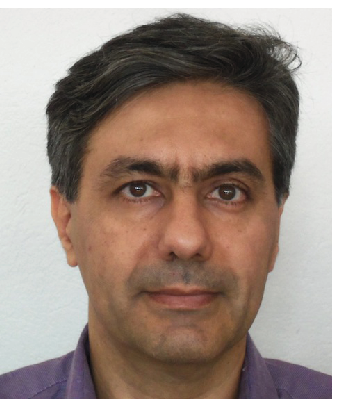}}]{Taufik Abr\~{a}o} (SM'12, SM-SBrT) received the B.S., M.Sc., and Ph.D. degrees in electrical engineering from the Polytechnic School of the University of S\~ao Paulo, S\~ao Paulo, Brazil, in 1992, 1996, and 2001, respectively. Since March 1997, he has been with the Communications Group, Department of Electrical Engineering, Londrina State University, Paran\'a, Brazil, where he is currently an Associate Professor in Telecommunications and the Head of the Telecomm. \& Signal Processing Lab. He is a Productivity Researcher from the CNPq Brazilian Agency. From July-October 2018 he was with the Connectivity section, Aalborg University as a Guest Researcher. In 2012, he was an Academic Visitor with the Southampton Wireless Research Group, University of Southampton, Southampton, U.K. From 2007 to 2008, he was a Post-doctoral Researcher with the Polytechnic University of Catalonia, Barcelona, Spain. He has served as Associate Editor for the IEEE Systems Journal (2020), the IEEE Access (2016-2018), IEEE Communication Surveys \& Tutorials (2013-2017), the AEUe-Elsevier (2020), the IET Signal Processing (2019), and JCIS-SBrT (2018-2020), and as Executive Editor of the ETT-Wiley (2016-2021) journal. His current research interests include communications and signal processing, especially massive MIMO, XL-MIMO, URLLC, mMTC,  optimization methods, machine learning, detection, estimation, resource allocation, and protocols.
\end{IEEEbiography}


\begin{IEEEbiography}[{\includegraphics[width=1in,height=1.25in,clip,keepaspectratio]{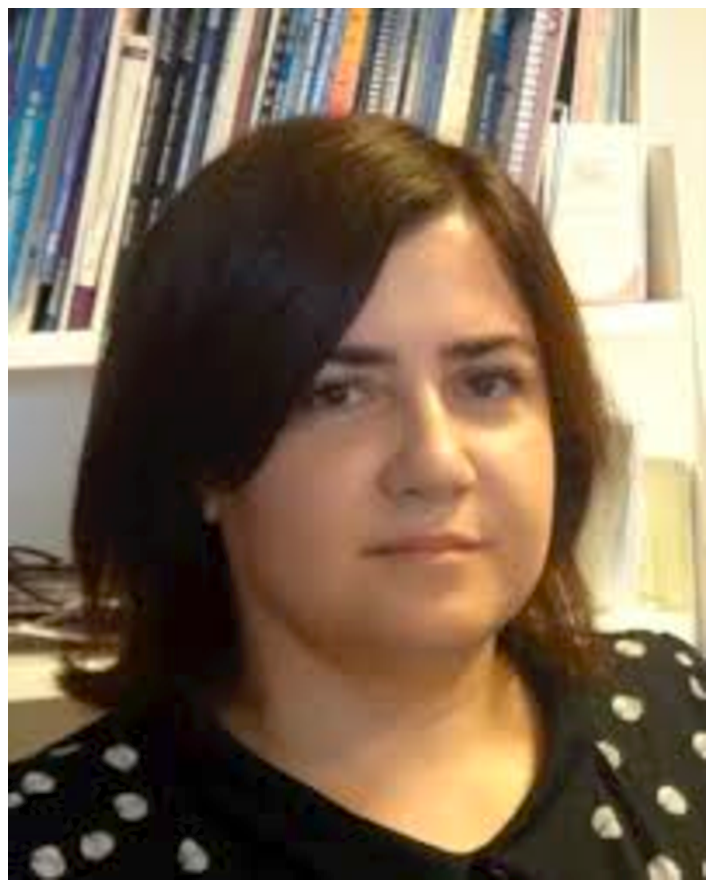}}]
{\bf Elisabeth de Carvalho}
received the Ph.D. degree in electrical engineering from Telecom ParisTech, France. She was a Post-Doctoral Fellow with Stanford University, Standford, CA, USA, and then with industry in the field of DSL and wireless LAN. Since 2005, she has been an Associate Professor with Aalborg University, where she has led several research projects in wireless communications. She co-authored the text book A Practical Guide to the MIMO Radio Channel. Her main expertise is in signal processing for MIMO communications, with recent focus on massive MIMO, including channel measurements, channel modeling, beamforming, and protocol aspects. 
\end{IEEEbiography}

\vspace{-70mm}

\begin{IEEEbiography}[{\includegraphics[width=1in,height=1.25in,clip,keepaspectratio]{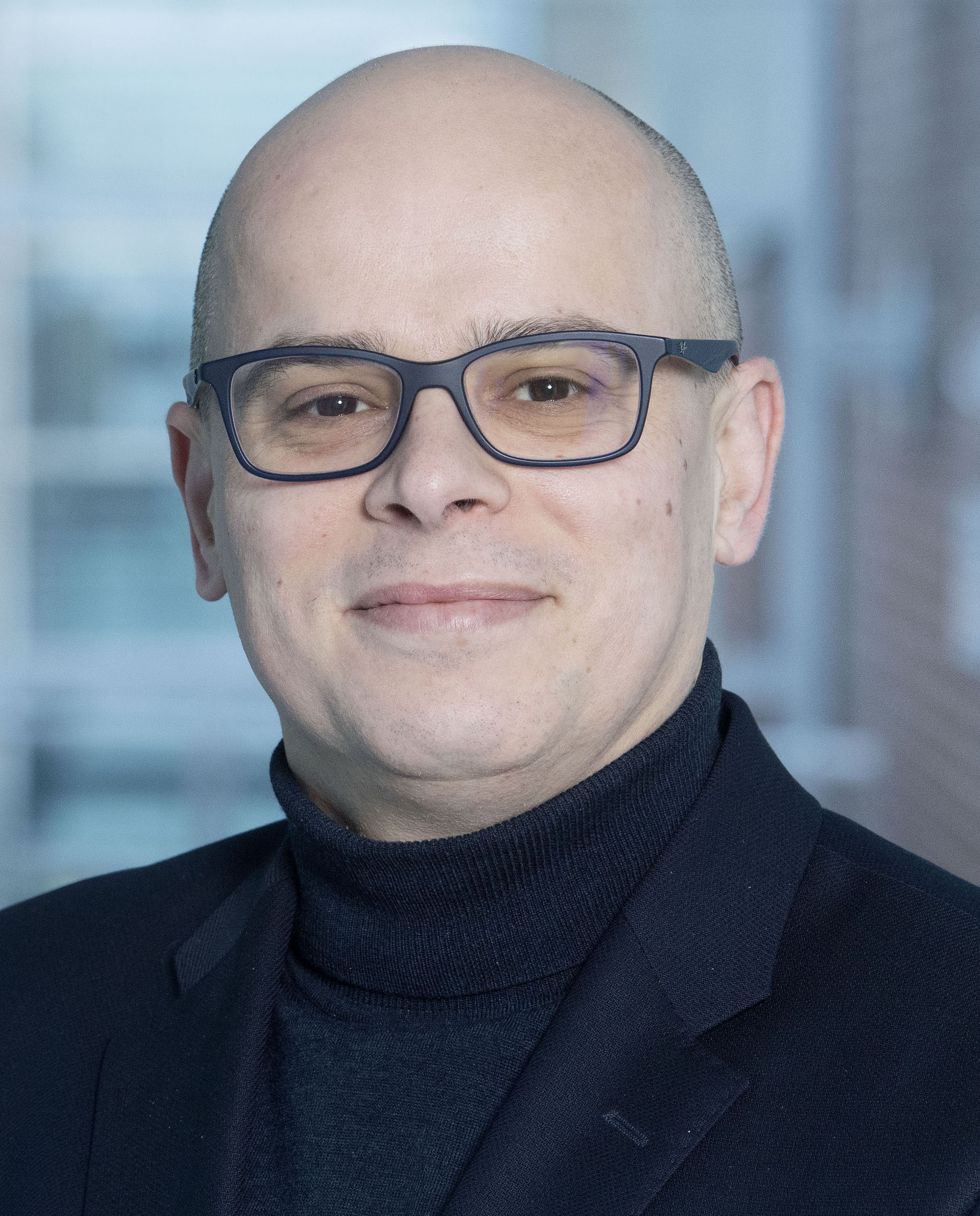}}]{\bf Petar Popovski}.
Petar Popovski (S'97--A'98--M'04--SM'10--F'16) is a Professor at Aalborg University, where he heads the section on Connectivity and a Visiting Excellence Chair at the University of Bremen. He received his Dipl.-Ing and M. Sc. degrees in communication engineering from the University of Sts. Cyril and Methodius in Skopje and the Ph.D. degree from Aalborg University in 2005. He is a Fellow of the IEEE. He received an ERC Consolidator Grant (2015), the Danish Elite Researcher award (2016), IEEE Fred W. Ellersick prize (2016), IEEE Stephen O. Rice prize (2018), Technical Achievement Award from the IEEE Technical Committee on Smart Grid Communications (2019), the Danish Telecommunication Prize (2020) and Villum Investigator Grant (2021). He is a Member at Large at the Board of Governors in IEEE Communication Society, Vice-Chair of the IEEE Communication Theory Technical Committee and IEEE TRANSACTIONS ON GREEN COMMUNICATIONS AND NETWORKING. He is currently an Area Editor of the IEEE TRANSACTIONS ON WIRELESS COMMUNICATIONS. Prof. Popovski was the General Chair for IEEE SmartGridComm 2018 and IEEE Communication Theory Workshop 2019. His research interests are in the area of wireless communication and communication theory. He authored the book ``Wireless Connectivity: An Intuitive and Fundamental Guide'', published by Wiley in 2020.
\end{IEEEbiography}


\end{document}

%% file: main.bbl

%% file: main.bbl
\begin{thebibliography}{10}
\providecommand{\url}[1]{#1}
\csname url@samestyle\endcsname
\providecommand{\newblock}{\relax}
\providecommand{\bibinfo}[2]{#2}
\providecommand{\BIBentrySTDinterwordspacing}{\spaceskip=0pt\relax}
\providecommand{\BIBentryALTinterwordstretchfactor}{4}
\providecommand{\BIBentryALTinterwordspacing}{\spaceskip=\fontdimen2\font plus
\BIBentryALTinterwordstretchfactor\fontdimen3\font minus
  \fontdimen4\font\relax}
\providecommand{\BIBforeignlanguage}[2]{{%
\expandafter\ifx\csname l@#1\endcsname\relax
\typeout{** WARNING: IEEEtran.bst: No hyphenation pattern has been}%
\typeout{** loaded for the language `#1'. Using the pattern for}%
\typeout{** the default language instead.}%
\else
\language=\csname l@#1\endcsname
\fi
#2}}
\providecommand{\BIBdecl}{\relax}
\BIBdecl

\bibitem{Bjornson2019}
E.~Bj{\"{o}}rnson, L.~Sanguinetti, H.~Wymeersch, J.~Hoydis, and T.~L. Marzetta,
  ``{Massive MIMO is a reality—What is next?: Five promising research
  directions for antenna arrays},'' \emph{Digital Signal Processing: A Review
  Journal}, vol.~94, pp. 3--20, 2019.

\bibitem{Carvalho2020}
E.~D. Carvalho, A.~Ali, A.~Amiri, M.~Angjelichinoski, and R.~W. Heath,
  ``{Non-stationarities in extra-large-scale massive MIMO},'' \emph{IEEE
  Wireless Communications}, vol.~27, no.~4, pp. 74--80, Aug. 2020.

\bibitem{Bjornson2017c}
E.~Bj{\"{o}}rnson, J.~Hoydis, and L.~Sanguinetti, ``{Massive MIMO networks:
  Spectral, energy, and hardware efficiency},'' \emph{Foundations and
  Trends{\textregistered} in Signal Processing}, vol.~11, no. 3-4, pp.
  154--655, 2017.

\bibitem{Ali2019a}
A.~Ali, E.~D. Carvalho, and R.~W. Heath, ``{Linear receivers in non-stationary
  massive MIMO channels with visibility regions},'' \emph{IEEE Wireless
  Communications Letters}, vol.~8, no.~3, pp. 885--888, Jun. 2019.

\bibitem{Sessler2005}
G.~M.~A. {Sessler} and F.~K. {Jondral}, ``{Low complexity polynomial expansion
  multiuser detector for CDMA systems},'' \emph{IEEE Transactions on Vehicular
  Technology}, vol.~54, no.~4, pp. 1379--1391, 2005.

\bibitem{Kammoun2014}
A.~{Kammoun}, A.~{Müller}, E.~{Björnson}, and M.~{Debbah}, ``{Linear
  precoding based on polynomial expansion: Large-scale multi-cell MIMO
  systems},'' \emph{IEEE Journal of Selected Topics in Signal Processing},
  vol.~8, no.~5, pp. 861--875, 2014.

\bibitem{Yin2014}
B.~Yin, M.~Wu, J.~R. Cavallaro, and C.~Studer, ``{Conjugate gradient-based
  soft-output detection and precoding in massive MIMO systems},'' in \emph{2014
  IEEE Global Communications Conference}.\hskip 1em plus 0.5em minus
  0.4em\relax IEEE, {Dec.} 2014, pp. 3696--3701.

\bibitem{Gao2014}
X.~Gao, L.~Dai, C.~Yuen, and Y.~Zhang, ``{Low-complexity MMSE signal detection
  based on Richardson method for large-scale MIMO systems},'' in \emph{2014
  IEEE 80th Vehicular Technology Conference (VTC2014-Fall)}.\hskip 1em plus
  0.5em minus 0.4em\relax IEEE, Sep. 2014, pp. 1--5.

\bibitem{Dai2015}
L.~Dai, X.~Gao, X.~Su, S.~Han, C.-L. I, and Z.~Wang, ``{Low-complexity
  soft-output signal detection based on Gauss–Seidel method for uplink
  multiuser large-scale MIMO systems},'' \emph{IEEE Transactions on Vehicular
  Technology}, vol.~64, no.~10, pp. 4839--4845, Oct. 2015.

\bibitem{Boroujerdi2018a}
M.~N. {Boroujerdi}, S.~{Haghighatshoar}, and G.~{Caire}, ``{Low-complexity
  statistically robust precoder/detector computation for massive MIMO
  systems},'' \emph{IEEE Transactions on Wireless Communications}, vol.~17,
  no.~10, pp. 6516--6530, Oct. 2018.

\bibitem{Kaczmarz1937}
S.~Kaczmarz, ``{Angen{\"{a}}herte aufl{\"{o}}sung von systemen linearer
  gleichungen},'' \emph{Bulletin International de l'Acad{\'{e}}mie Polonaise
  des Sciences et des Lettres. Classe des Sciences Math{\'{e}}matiques et
  Naturelles. S{\'{e}}rie A, Sciences Math{\'{e}}matiques}, vol.~35, pp.
  355--357, 1937.

\bibitem{Strohmer2009a}
T.~Strohmer and R.~Vershynin, ``{A randomized Kaczmarz algorithm with
  exponential convergence},'' \emph{Journal of Fourier Analysis and
  Applications}, vol.~15, no.~2, pp. 262--278, Apr. 2009.

\bibitem{Bai2018a}
Z.~Z. Bai and W.~T. Wu, ``{On greedy randomized Kaczmarz method for solving
  large sparse linear systems},'' \emph{SIAM Journal on Scientific Computing},
  vol.~40, no.~1, pp. A592--A606, 2018.

\bibitem{Censor2009}
Y.~Censor, G.~T. Herman, and M.~Jiang, ``{A note on the behavior of the
  randomized Kaczmarz algorithm of Strohmer and Vershynin},'' \emph{Journal of
  Fourier Analysis and Applications}, vol.~15, no.~4, pp. 431--436, 2009.

\bibitem{Strohmer2009}
T.~Strohmer and R.~Vershynin, ``{Comments on the randomized Kaczmarz method},''
  \emph{Journal of Fourier Analysis and Applications}, vol.~15, no.~4, pp.
  437--440, 2009.

\bibitem{Sun2020}
M.-L. Sun, C.-Q. Gu, and P.-F. Tang, ``{On randomized sampling Kaczmarz method
  with application in compressed sensing},'' \emph{Mathematical Problems in
  Engineering}, vol. 2020, pp. 1--11, Mar. 2020.

\bibitem{Amiri2020}
A.~Amiri, S.~Rezaie, C.~N. Manchon, and E.~de~Carvalho, ``{Distributed
  receivers for extra-large scale MIMO arrays: A message passing approach},''
  \emph{{arXiv}}, pp. 1--31, 2020.

\bibitem{Wang2020}
\BIBentryALTinterwordspacing
H.~Wang, A.~Kosasih, C.-K. Wen, S.~Jin, and W.~Hardjawana, ``{Expectation
  propagation detector for extra-large scale massive MIMO},'' \emph{IEEE
  Transactions on Wireless Communications}, vol.~19, no.~3, p. 2036–2051,
  Mar. 2020. [Online]. Available:
  \url{http://dx.doi.org/10.1109/TWC.2019.2961892}
\BIBentrySTDinterwordspacing

\bibitem{Rodrigues2020}
V.~C. Rodrigues, A.~Amiri, T.~Abrao, E.~de~Carvalho, and P.~Popovski,
  ``{Low-complexity distributed XL-MIMO for multiuser detection},'' in
  \emph{2020 IEEE International Conference on Communications Workshops (ICC
  Workshops)}.\hskip 1em plus 0.5em minus 0.4em\relax IEEE, Jun. 2020, pp.
  1--6.

\bibitem{Meyer2000}
C.~D. Meyer, \emph{{Matrix analysis and applied linear algebra}}.\hskip 1em
  plus 0.5em minus 0.4em\relax SIAM, 2000.

\bibitem{Wu2020}
{Wu, Hebiao and Shen, Bin and Zhao, Shufeng and Gong, Peng}, ``{Low-complexity
  soft-output signal detection based on improved Kaczmarz iteration algorithm
  for Uplink massive MIMO system},'' \emph{Sensors}, vol.~20, no.~6, p. 1564,
  Mar. 2020.

\bibitem{Press2007}
W.~H. Press, S.~A. Teukolsky, W.~T. Vetterling, and B.~P. Flannery,
  \emph{{Numerical recipes: The art of scientific computing}}, 3rd~ed.\hskip
  1em plus 0.5em minus 0.4em\relax Cambridge, UK: Cambridge University Press,
  2007.

\bibitem{Bai2019}
Z.~Z. Bai and W.~T. Wu, ``{On partially randomized extended Kaczmarz method for
  solving large sparse overdetermined inconsistent linear systems},''
  \emph{Linear Algebra and Its Applications}, vol. 578, pp. 225--250, 2019.

\bibitem{Rodrigues2019}
V.~C. Rodrigues, J.~C. {Marinello Filho}, and T.~Abr{\~{a}}o, ``{Randomized
  Kaczmarz algorithm for Massive MIMO systems with channel estimation and
  spatial correlation},'' \emph{International Journal of Communication
  Systems}, p. e4158, Sep. 2019.

\bibitem{Bai2018}
Z.~Z. Bai and W.~T. Wu, ``{On relaxed greedy randomized Kaczmarz methods for
  solving large sparse linear systems},'' \emph{Applied Mathematics Letters},
  vol.~83, pp. 21--26, 2018.

\bibitem{Carvalho2012}
T.~Brown, E.~{De Carvalho}, and P.~Kyritsi,
  \emph{\BIBforeignlanguage{English}{Practical guide to the MIMO radio channel
  with MATLAB examples}}, 1st~ed.\hskip 1em plus 0.5em minus 0.4em\relax Wiley,
  2012.

\bibitem{Bjornson2018}
E.~Bjornson, J.~Hoydis, and L.~Sanguinetti, ``{Massive MIMO has unlimited
  capacity},'' \emph{IEEE Transactions on Wireless Communications}, vol.~17,
  no.~1, pp. 574--590, Jan. 2018.

\end{thebibliography}
